\documentclass[graybox]{svmult}

\usepackage{type1cm}        
\usepackage{makeidx}         
\usepackage{graphicx}        
\usepackage{multicol}        
\usepackage[bottom]{footmisc}

\usepackage{newtxtext}       %
\usepackage{newtxmath}       
\usepackage{algorithm}
\usepackage{algorithmicx}
\usepackage{algpseudocode}
\usepackage{wrapfig}
\usepackage{rotating} 

\usepackage{color}
\usepackage{tikz}
\usepackage{pgfplots}
\pgfplotsset{compat=1.10}
\usetikzlibrary{patterns}
\usepackage{pgfplotstable}
\usepackage{csvsimple}
\definecolor{excelblue}{RGB}{94,156,211}
\definecolor{excelorange}{RGB}{235,125,60}
\definecolor{excelgray}{RGB}{165,165,165}
\definecolor{excelgreen}{RGB}{114,172,77}

\definecolor{mycolor0}{HTML}{2B83BA}
\definecolor{mycolor1}{HTML}{D7301F}
\definecolor{mycolor2}{HTML}{FC8D59}
\definecolor{mycolor3}{HTML}{abdda4}
\definecolor{mycolor4}{HTML}{e9a3c9}
\definecolor{mycolor5}{HTML}{FEF0D9}

\usepackage{newtxtext}       %
\usepackage{newtxmath}       

\usepackage{listings}
\makeindex             

\newcommand{\ml}       {\textsc{Matlab}}

\newcommand{\nl}{\scriptstyle{0}}
\newcommand{\mmd}       {\textsc{Mmd}}
\newcommand{\amd}       {\textsc{Amd}}

\newcommand{\cM}{\mathcal{M}}

\newcommand{\rcm}       {\textsc{Rcm}}
\newcommand{\metis}     {\textsc{Metis}}
\newcommand{\parmetis}     {\textsc{ParMetis}}
\newcommand{\mtmetis}     {\textsc{MT-Metis}}
\newcommand{\scotch}     {\textsc{Scotch}}
\newcommand{\mat}[1]{\left(\begin{array}{#1}}
\newcommand{\rix}{\end{array}\right)}
\newcommand{\R}{\mathbb{R}}
\newcommand{\circn}[1]{{\Large$\bigcirc\mkern-12.0mu$}$#1\;$}

\newcommand{\nxlnz}{xl}
\newcommand{\nlnz}{l}
\newcommand{\nindx}{id}
\newcommand{\nxindx}{xid}
\newcommand{\nr}{r}
\newcommand{\ntemp}{t}

\newcommand{\vxlnz}{\texttt{\nxlnz{}}}
\newcommand{\vlnz}{\texttt{\nlnz}}
\newcommand{\vxindx}{\texttt{\nxindx}}
\newcommand{\vindx}{\texttt{\nindx}}
\newcommand{\vr}{\texttt{\nr}}
\newcommand{\vtemp}{\texttt{\ntemp}}

\newcommand{\bi}{\begin{itemize}}
\newcommand{\ei}{\end{itemize}}

\newcommand{\be}{\begin{equation}}
\newcommand{\ee}{\end{equation}}

\begin{document}

\title*{State-of-The-Art Sparse Direct Solvers}
\author{Matthias Bollh\"ofer \and Olaf Schenk \and Radim Janal\'ik \and Steve Hamm \and Kiran Gullapalli}
\authorrunning{M. Bollh\"ofer, O. Schenk, R. Janal\'ik, Steve Hamm, and  K. Gullapalli}
\institute{Matthias Bollh\"ofer \at Institute for Computational Mathematics, TU Braunschweig, Germany, \email{m.bollhoefer@tu-bs.de}
\and 
Olaf Schenk \at Institute of Computational Science, Faculty of Informatics, Universit\`a della Svizzera italiana, Switzerland, \email{olaf.schenk@usi.ch}
\and  
Radim Janal\'ik \at Institute of Computational Science, Faculty of Informatics, Universit\`a della Svizzera italiana, Switzerland, \email{radim.janalik@usi.ch}
\and 
Steve Hamm \at NXP, United States of America, \email{steve.hamm@nxp.com}
\and
Kiran Gullapalli \at NXP, United States of America, \email{kiran.gullapalli@nxp.com}}
%
%
\maketitle

\abstract*{In this chapter we will give an insight into modern sparse elimination
methods. These are driven by a preprocessing phase based on
combinatorial algorithms which improve diagonal dominance, reduce fill-in, and
improve concurrency to allow for parallel treatment. Moreover, these methods
detect dense submatrices which can be handled by dense matrix kernels based on
multithreaded level-3 BLAS. We will demonstrate for problems arising from
circuit simulation, how the improvements in recent years have advanced 
direct solution methods significantly.
}
\abstract{In this chapter we will give an insight into modern sparse elimination
methods. These are driven by a preprocessing phase based on
combinatorial algorithms which improve diagonal dominance, reduce fill--in and
improve concurrency to allow for parallel treatment. Moreover, these methods
detect dense submatrices which can be handled by dense matrix kernels based on
multi-threaded level--3 BLAS. We will demonstrate for problems arising from
circuit simulation how the improvement in recent years have advanced 
direct solution methods significantly.
}

\section{Introduction}
\label{sec:intro}
Solving large sparse linear systems is at the heart of many application
problems arising from engineering problems. Advances in combinatorial
methods in combination with modern computer architectures have massively
influenced the design of state-of-the-art direct solvers
that are feasible for solving  larger systems efficiently
in a computational environment with rapidly increasing memory resources
and cores. Among these advances are 
novel combinatorial algorithms for improving diagonal dominance which
pave the way to a static pivoting approach, thus improving the 
efficiency of the factorization
phase dramatically. Besides, partitioning and reordering the system
such that a high level of concurrency is achieved, the objective is to 
simultaneously achieve the reduction of fill-in and the parallel concurrency.
While these achievements already significantly improve the factorization
phase, modern computer architectures require one to compute as many operations
as possible in the cache of the CPU. This in turn can be achieved when
dense subblocks that show up during the factorization can be grouped
together into dense submatrices which are handled
by multithreaded and cache-optimized 
dense matrix kernels using level-3 BLAS and LAPACK
\cite{AndBBDDDGHMOS95}.

This chapter will review some of the basic technologies together
with the latest developments for sparse direct solution methods that have led to
state-of-the-art $LU$ decomposition methods.
The paper is organized as follows. In Section \ref{sec:mwm}
we will start with maximum weighted matchings which is one of the key
tools in combinatorial optimization to dramatically improve the diagonal dominance
of the underlying system.
Next, Section \ref{sec:reordering} will review multilevel nested dissection
as a combinatorial method to reorder a system symmetrically
such that fill-in and parallelization can are improved simultaneously, once
pivoting can be more or less ignored.
After that, we will review established graph-theoretical approaches
in Section \ref{sec:lu}, in particular the elimination tree, from which
most of the properties of the $LU$ factorization can be concluded. Among
these properties is the prediction of dense submatrices in the
factorization. In this way several subsequent
columns of the factors $L$ and $U^T$ are collected in a single dense block. 
This is the basis for the use of dense matrix kernels using optimized
level-3 BLAS as well to exploit fast computation using the cache hierarchy which 
is discussed in Section~\ref{sec:parallel}.
Finally we will demonstrate in Section~\ref{sec:appl}, for
examples from circuit simulation, how the ongoing developments in 
sparse direct solution methods have accelerated state-of-the-art solution 
techniques.
We assume that the reader is familiar with some elementary knowledge from
graph theory, see; e.g., \cite{DufER86,GeoL81} and some simple
computational algorithms based on graphs \cite{AhoHU83}.

\section{Maximum weight matching}
\label{sec:mwm}
\begin{figure}
\begin{minipage}{.33\textwidth}
 \begin{center}
 Original Matrix $A$

    $\left(
        \begin{array}{cccccc}
         1 & 3 & \nl &  2 & \nl & \nl \\
         3  & \nl & \nl & 4 & \nl & 1 \\
        \nl & \nl & \nl & \nl &  3  & \nl \\
        2 &  4  & \nl & \nl & 1 & \nl \\
        \nl & \nl &  3  & 1 & \nl & \nl \\
        \nl & 1 & \nl & \nl & \nl &  2
        \end{array}
    \right) $ 
 \end{center}
\end{minipage}
\begin{minipage}{.32\textwidth}
 \begin{center}
$G_b(A): \;$ 
\includegraphics[width=0.43\textwidth]{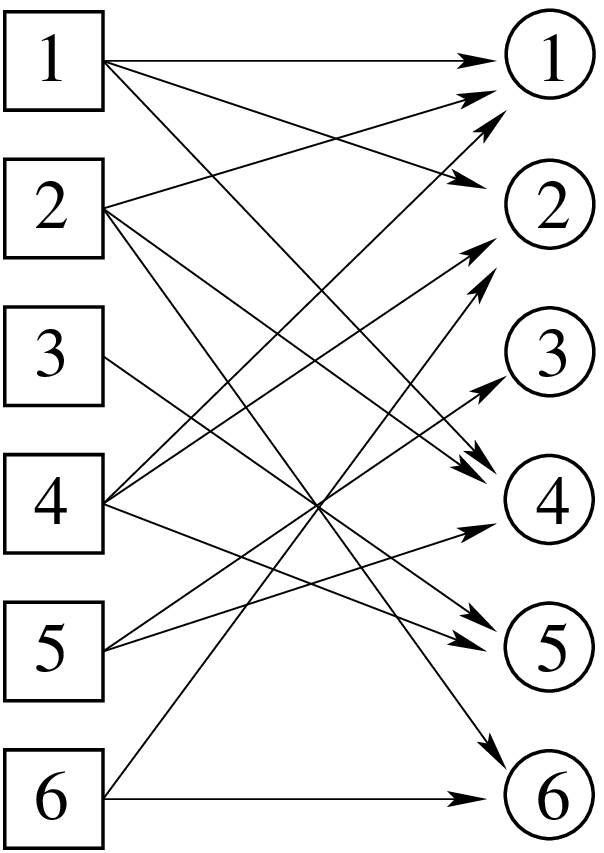} 
%

\hspace{0.5cm}$\cM: \;$ 
\includegraphics[width=0.43\textwidth]{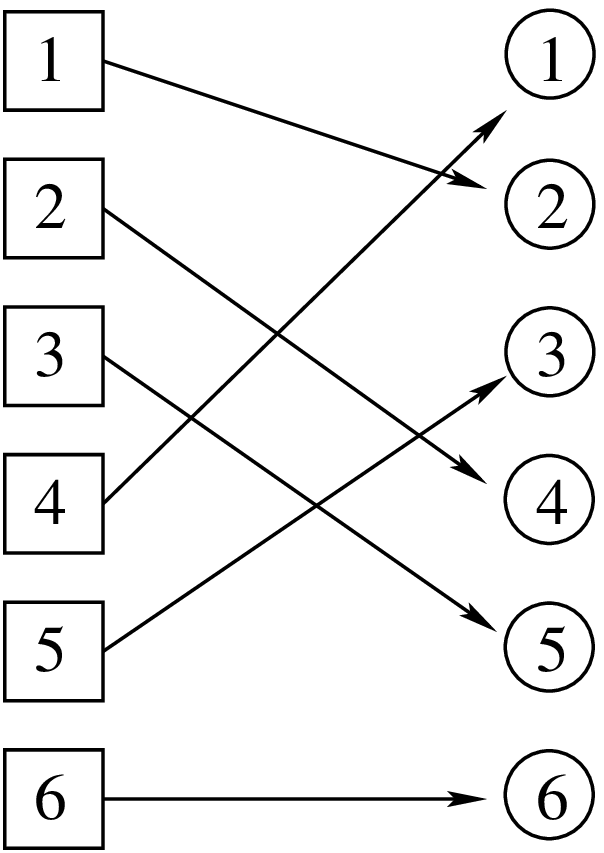} 
 \end{center}
\end{minipage}
\begin{minipage}{.32\textwidth}
  \begin{center}
Reordered Matrix $\Pi^TA$

\medskip

 \end{center}  
\end{minipage}
    \caption{Perfect matching. Left side: original
      matrix $A$. Middle: bipartite representation $G_b(A) = (V_r, V_c, E)$
      of the matrix $A$ and perfect matching $\cM$. Right side: permuted matrix
      $\Pi^TA$.}
    \label{fig:unsym_perm}
\end{figure}

In modern sparse elimination methods the key to success 
is ability to work with efficient data structures and their underlying
numerical templates. If we can increase the size of the diagonal entries
as much as possible in advance, pivoting during Gaussian elimination can often 
be bypassed and we may work with static data structures and
the numerical method will be significantly accelerated. 
A popular method to achieve this goal is the
maximum weight matching method~\cite{DufK99S,olschowka:1996} 
which permutes (e.g.) the rows of a given
nonsingular matrix $A\in\R^{n,n}$ by a permutation matrix $\Pi\in\R^{n,n}$ 
such that $\Pi^TA$
has a \emph{non-zero diagonal}. Moreover, it maximizes 
the product of the absolute diagonal values  and yields diagonal
scaling matrices $D_r, D_c\in\R^{n,n}$ such that $\tilde A=\Pi^TD_rAD_c$ satisfies
$|\tilde a_{ij}|\leqslant 1$ and $|\tilde a_{ii}|=1$ for all $i,j=1,\dots,n$.
The original idea on which these nonsymmetric permutations and scalings are
based is to find a \emph{maximum weighted matching} of a
\emph{bipartite graphs}. Finding a maximum weighted matching is a well
known assignment problem in operation research and combinatorial
analysis.
\begin{definition}
A graph $G=(V,E)$ with vertices $V$ and edges $E\subset V^2$ 
is called \emph{bipartite} 
if $V$ can be partitioned into two sets  $V_r$ and  $V_c$, such that no edge
$e=(v_1,v_2) \in E$ has both ends $v_1,v_2$ in $V_r$ or both ends $v_1,v_2$ 
in $V_c$. In this case we denote $G$ by $G_b=(V_r,V_c,E)$.
\end{definition}
\begin{definition}
Given a matrix $A$, then we can associate with it a canonical
bipartite graph $G_b(A)=(V_r,V_c,E)$ by assigning the 
labels of $V_r=\{r_1,\dots,r_n\}$ 
with the row indices of $A$ and 
$V_c=\{c_1,\dots,c_n\}$ being labeled by the column indices.
In this case $E$ is defined via $E=\{(r_i,c_j)|\; a_{ij}\not=0\}$.
\end{definition}
For the bipartite graph $G_b(A)$ we see immediately that 
If $a_{ij}\not=0$, 
then we have that $r_i \in V_r$ from the row set
is connected by an edge $(r_i,c_j) \in E$ to the column $c_j \in V_c$,
but neither rows are connected with each other nor do the columns have
inter connections.
\begin{definition}\label{def:matching}
A \emph{matching} $\cM$ 
of a given graph $G= (V,E)$ is a subset of edges
$e\in E$ such that no two of which share the same vertex. 
\end{definition}
If $\cM$ is a
matching of a bipartite graph $G_b(A)$, then each edge $e=(r_i,c_j) \in \cM$ 
corresponds to a row $i$ and a column $j$ and there exists no other edge 
$\hat e=(r_k,c_l) \in \cM$ 
that has the same vertices, neither $r_k=r_i$ nor $c_l=c_j$. 
\begin{definition}\label{def:maxmatching}
A matching $\cM$ of $G=(V,E)$ is called
\emph{maximal}, if no other edge from $E$ can be added to $\cM$.
\end{definition}
If, for an $n \times n$ matrix $A$ a \emph{matching} $\cM$ of $G_b(A)$ with
maximum cardinality $n$ is found, then by definition the edges 
must be $(i_1,1),\dots,(i_n,n)$ with $i_1,\dots,i_n$ being the 
numbers $1,\dots,n$ in a suitable order and therefore we obtain
$a_{i_1,1}\not=0$, \dots
$a_{i_n,n}\not=0$. In this case 
we have established that the
matrix $A$ is at least structurally nonsingular and we can use a 
row permutation matrix $\Pi^T$ associated with row ordering $i_1,\dots,i_n$ 
to place a nonzero entry on each diagonal location of $\Pi^TA$.
\begin{definition}
A \emph{perfect matching} is a maximal matching with cardinality $n$.
\end{definition}
It can be shown that for a structurally nonsingular matrix $A$ there always
exists a perfect matching $\cM$.
\begin{example}{Perfect Matching}
In Figure \ref{fig:unsym_perm}, the set of edges $\cM= \{(1,2), (2,4),
(3,5), (4,1), (5,3), (6,6) \}$ represents a perfect maximum matching
of the bipartite graph $G_b(A)$.
\end{example}

The most efficient combinatorial methods for finding maximum matchings
in bipartite graphs make use of an \emph{augmenting path}. We will
introduce some graph terminology for the
 construction of perfect
matchings. 
\begin{definition}
If an edge $e=(u,v)$ in a graph $G=(V,E)$
joins a vertices $u,v\in V$, then we denote it as $uv$. 
A path then consists of edges $u_1u_2,u_2u_3,u_3u_4 \ldots,u_{k-1}u_k$, where 
each $(u_i,u_{i+1})\in E$, $i=1,\dots,k-1$.
\end{definition}
If $G_b=(V_r,V_c,E)$ is a bipartite graph, then by definition of a path, 
any path is alternating between the vertices of $V_r$ and $V_c$, e.g.,
pathes in $G_b$ could be such as $r_1c_2,c_2r_3,r_3c_4,\dots$.
\begin{definition}
Given a graph $G=(V,E)$, a 
vertex is called \emph{free} if it is not
incident to any other edge in a matching $\cM$ of $G$. 
An \emph{alternating path} relative to a matching $\cM$ is a path 
$P = u_1u_2,u_2u_3, \ldots,u_{s-1}u_s$ where its edges are alternating 
between $E \setminus \cM$ and $\cM$. An
\emph{augmenting path} relative to a matching $\cM$ is an alternating
path of odd length and both of it vertex endpoints are free. 
\end{definition}
\begin{example}{Augmenting Path}
Consider Figure \ref{fig:unsym_perm}.
To better distinguish between row and column vertices we use
$\fbox{$1$},\fbox{$2$},\dots,\fbox{$6$}$ for the rows and \circn{1},\circn{2},\dots,\circn{6} for the
columns.
A non-perfect but maximal matching is given by
$M= \{(\fbox{$4$},$\circn{5}$), (\fbox{$1$},$\circn{1}$), (\fbox{$6$},$\circn{2}$), (\fbox{$2$},$\circn{6}$), (\fbox{$5$},$\circn{4}$) \}$. 
We can easily see that an augmenting path 
alternating between rows and columns is given by \fbox{$3$}\circn{5} , \circn{5}\fbox{$4$} , \fbox{$4$}\circn{1} , \circn{1}\fbox{$1$} , \fbox{$1$}\circn{2} , \circn{2}\fbox{$6$} , \fbox{$6$}\circn{6} , \circn{6}\fbox{$2$} , \fbox{$2$}\circn{4} , \circn{4}\fbox{$5$} , \fbox{$5$}\circn{3}. Both endpoints \fbox{$3$} and \circn{3}
of this augmenting path are free.
\end{example}

In a bipartite
graph $G_b= (V_r, V_c, E)$ one vertex endpoint of any
augmenting path must be in $V_r$ whereas the other one must be in $V_c$. 
The symmetric
difference, $A \oplus B$ of two edge sets $A$, $B$ is defined to be $(A
\setminus B) \cup (B \setminus A)$.

Using these definitions and notations,
the following theorem \cite{Berge} gives an
constructive algorithm for finding perfect matchings in bipartite
graphs.

\begin{theorem}\label{theo:Berge}
If $\cM$ is non-maximum matching of a bipartite graph $G_b= (V_r, V_c,E)$, 
then there exists an augmenting path $P$ relative to $\cM$ such that
 $P=\tilde{\cM} \oplus \cM$ and $\tilde{\cM}$
is a matching with cardinality $|\cM|+1$.
\end{theorem}
According to this theorem, a combinatorial method of finding perfect
matching in a bipartite graph is to seek for augmenting paths. 

The
perfect matching as discussed so far only takes the nonzero structure
of the matrix into account. 
For their use as static pivoting methods prior to the $LU$ decomposition
one requires in addition to
maximize the absolute value of the product of the diagonal entries. 
This is referred to as
\emph{maximum weighted matching}. In this case a permutation
$\pi$ has to be found, which maximizes
\begin{equation}
  \prod_{i=1}^n |a_{\pi(i)i}|. \label{eq:1}
\end{equation}
The maximization of this product is transferred into a minimization of a sum as follows. We define a matrix $C = (c_{ij})$ via
\[
  c_{ij} = 
  \begin{cases}
    \log a_i - \log |a_{ij}| & a_{ij} \neq 0 \\
    \infty                     & \text{otherwise},
  \end{cases}
\]
where $a_i = \max_j |a_{ij}|$  is the maximum element in row $i$ of
matrix $A$. A permutation $\pi$ which minimizes the sum 
\[
  \label{eq:4}
  \sum_{i=1}^n c_{\pi(i)i} 
\]
also maximizes the product~(\ref{eq:1}). The minimization problem is
known as linear-sum assignment problem or bipartite weighted matching
problem in combinatorial optimization.  The problem is solved by a
sparse variant of the Hungarian method. The complexity is
 $\mathcal{O}(n \tau \log n )$ for
sparse matrices with $\tau$ entries. For matrices, whose associated
graph fulfill special requirements, this bound can be reduced further
to $\mathcal{O}(n^\alpha (\tau + n \log n)$ with $\alpha < 1$.  All graphs
arising from finite-difference or finite element discretizations meet
the conditions~\cite{gupta:99}. As before, we finally get a perfect
matching which in turn defines a nonsymmetric permutation.

When solving the assignment problem, two dual vectors $u = (u_i)$ and
$v = (v_i)$ are computed which satisfy
\begin{align}
  u_i + v_j & = c_{ij} \qquad  (i,j) \in \cM, \label{eq:11} \\
  u_i + v_j & \leq c_{ij} \qquad \text{otherwise}. \label{eq:12}
\end{align}
Using the exponential function these vectors can be used to scale the 
initial matrix. To do so define
two diagonal matrices $D_r$ and $D_c$ through
\begin{align}
  D_r & = \text{diag}(d_1^r,d_2^r,\dots,d_n^r), \qquad d_i^r = \exp(u_i),\\ 
  D_c & = \text{diag}(d_1^c,d_2^c,\dots,d_n^c), \qquad d_j^c = \exp(v_j)/a_j.
\end{align}
Using equations (\ref{eq:11}) and (\ref{eq:12}) and the definition of $C$, 
it immediately follows that $\tilde A = \Pi^T D_r A D_c$ satisfies
\begin{align}
  |\tilde a_{ii}| & = 1, \label{eq:13}\\
  |\tilde a_{ij}| & \le 1. \label{eq:14}
\end{align}
The permuted and scaled system $\tilde A$ has been observed to
have significantly better numerical properties when being used
for direct methods or for preconditioned iterative methods, cf. e.g. 
\cite{benzi:2000:phi,DufK99S}. Olschowka and
Neumaier~\cite{olschowka:1996} introduced these scalings and
permutation for reducing pivoting in Gaussian elimination of full
matrices. The first implementation for sparse matrix problems was
introduced by Duff and Koster~\cite{DufK99S}.  For symmetric
matrices $|A|$, these nonsymmetric matchings can be converted
to a symmetric permutation $P$ and a symmetric scaling $D_s=(D_rD_c)^{1/2}$
such that $P^TD_sAD_sP$ consists mostly of diagonal blocks of size $1\times 1$
and $2\times 2$ satisfying a similar condition as (\ref{eq:13}) and (\ref{eq:14}),
where in practice it rarely happens that  $1\times 1$ blocks are identical 
to $0$~\cite{dupr:04a}.
Recently, successful parallel approaches to compute maximum weighted matchings have
been proposed~\cite{LanPM11,LanAM14}.

\begin{figure}
\begin{minipage}{.45\textwidth}
 \begin{center}
%
\includegraphics[width=0.8\textwidth]{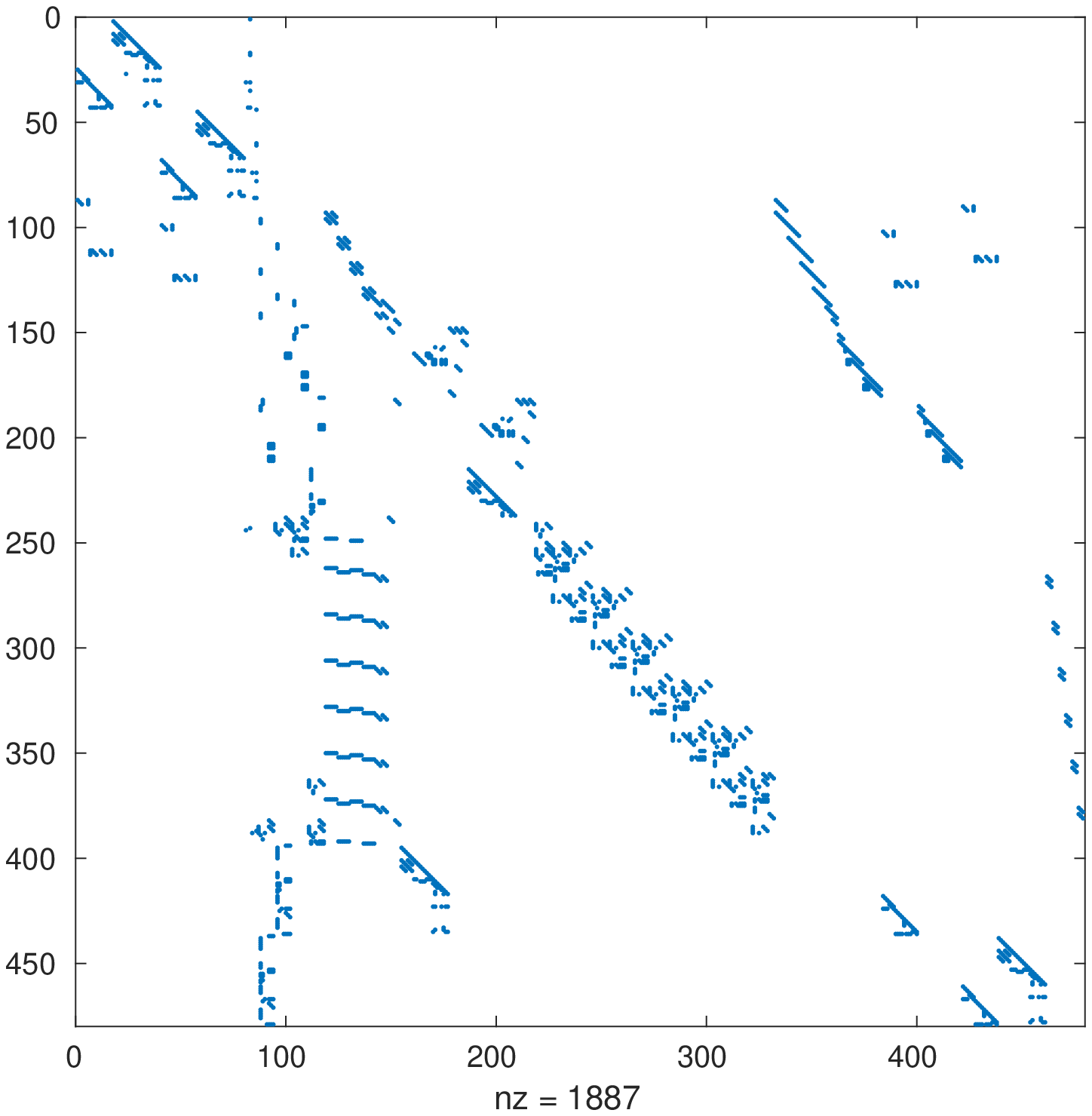} 
 \end{center}
\end{minipage}
~
\begin{minipage}{.45\textwidth}
  \begin{center}
%
\includegraphics[width=0.8\textwidth]{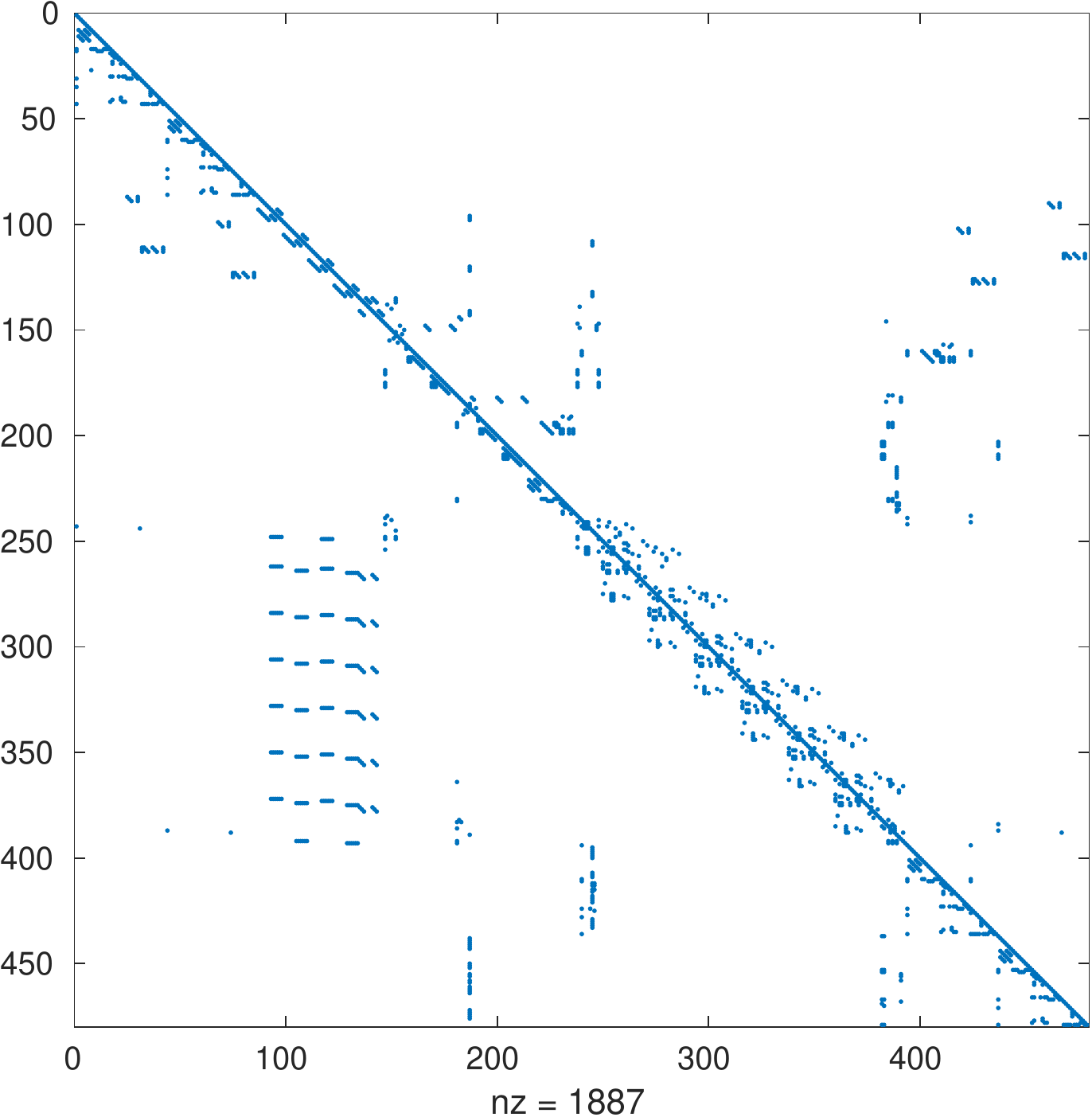} 
 \end{center}  
\end{minipage}
    \caption{Maximum weight matching. Left side: original
      matrix $A$. Right side: permuted and rescaled matrix
      $\tilde A=\Pi^TD_rAD_c$.}
    \label{fig:mwm}
\end{figure}

\begin{example}{Maximum Weight Matching}\label{exm:west0479-match}
To conclude this section we demonstrate the effectiveness of maximum weight matchings using a simple sample matrix ``west0479'' from the SuiteSparse Matrix Collection.
The matrix can also directly be loaded in \ml{} using \texttt{load west0479}.
In Figure \ref{fig:mwm} we display the matrix before and after applying maximum
weighted matchings. To illustrate the improved diagonal dominance we further
compute $r_i=|a_{ii}|/\sum_{j=1}^n|a_{ij}|$ for each row of $A$ and $\tilde A=\Pi^TD_rAD_s$, $i=1,\dots,n$. $r_i$ can be read as relative diagonal dominance of row $i$ 
and yields a number between $0$ and $1$. Moreover, whenever $r_i>\frac12$, the row
is strictly diagonal dominant, i.e., $|a_{ii}|>\sum_{j:j\not=i}|a_{ij}|$.
In Figure \ref{fig:mwm} we display for both matrices $r_i$ by sorting its values
in increasing order and taking $\frac12$ as reference line. We can see the
dramatic impact of maximum weighted matchings in improving the diagonal dominance
of the given matrix and thus paving its way to a static pivoting approach
in incomplete or complete $LU$ decomposition methods.
\end{example}
\begin{figure}
\begin{minipage}{.48\textwidth}
 \begin{center}
%
\includegraphics[width=0.95\textwidth,height=0.5\textwidth]{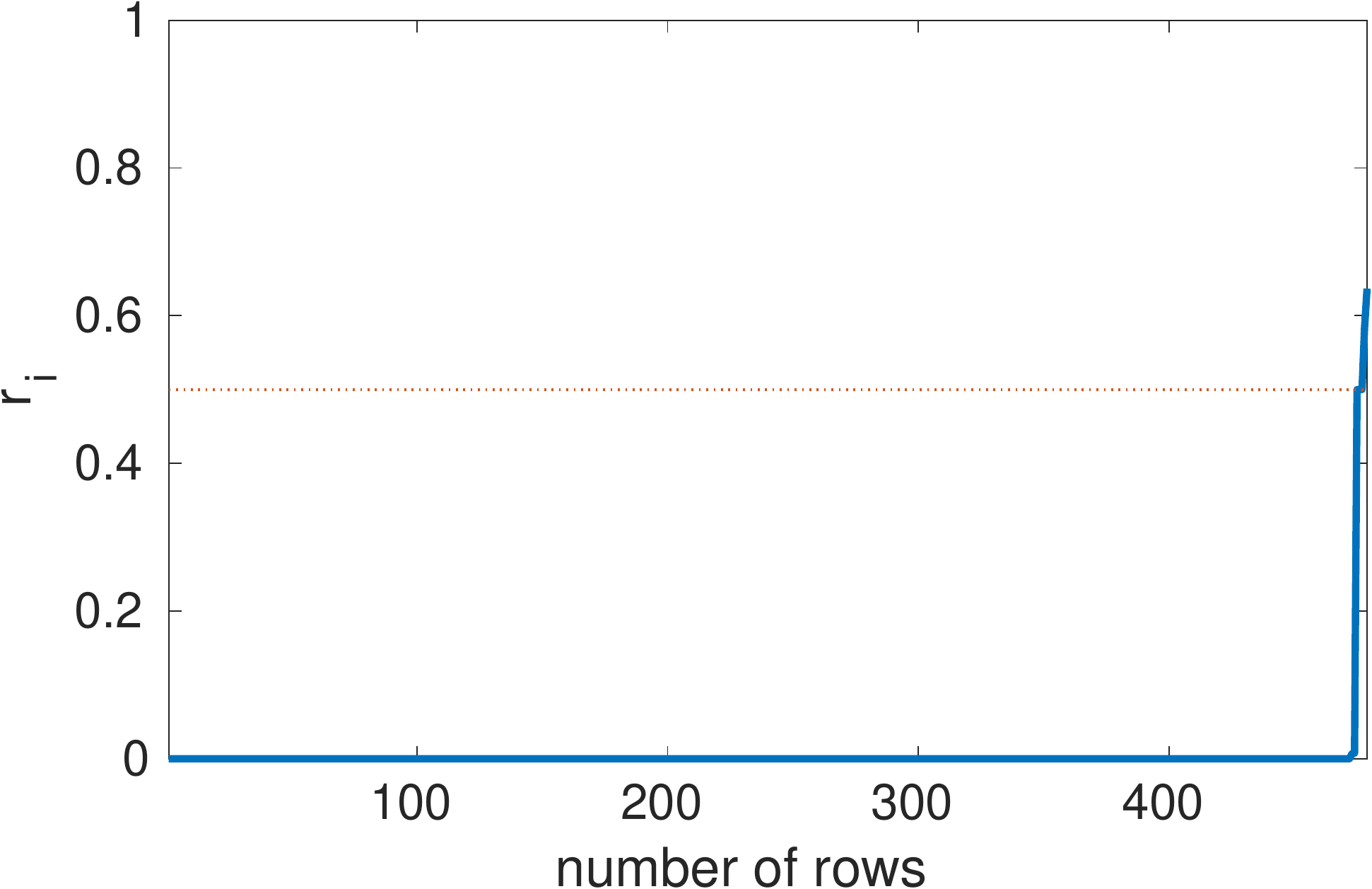} 
 \end{center}
\end{minipage}
~
\begin{minipage}{.48\textwidth}
  \begin{center}
%
\includegraphics[width=0.95\textwidth,height=0.5\textwidth]{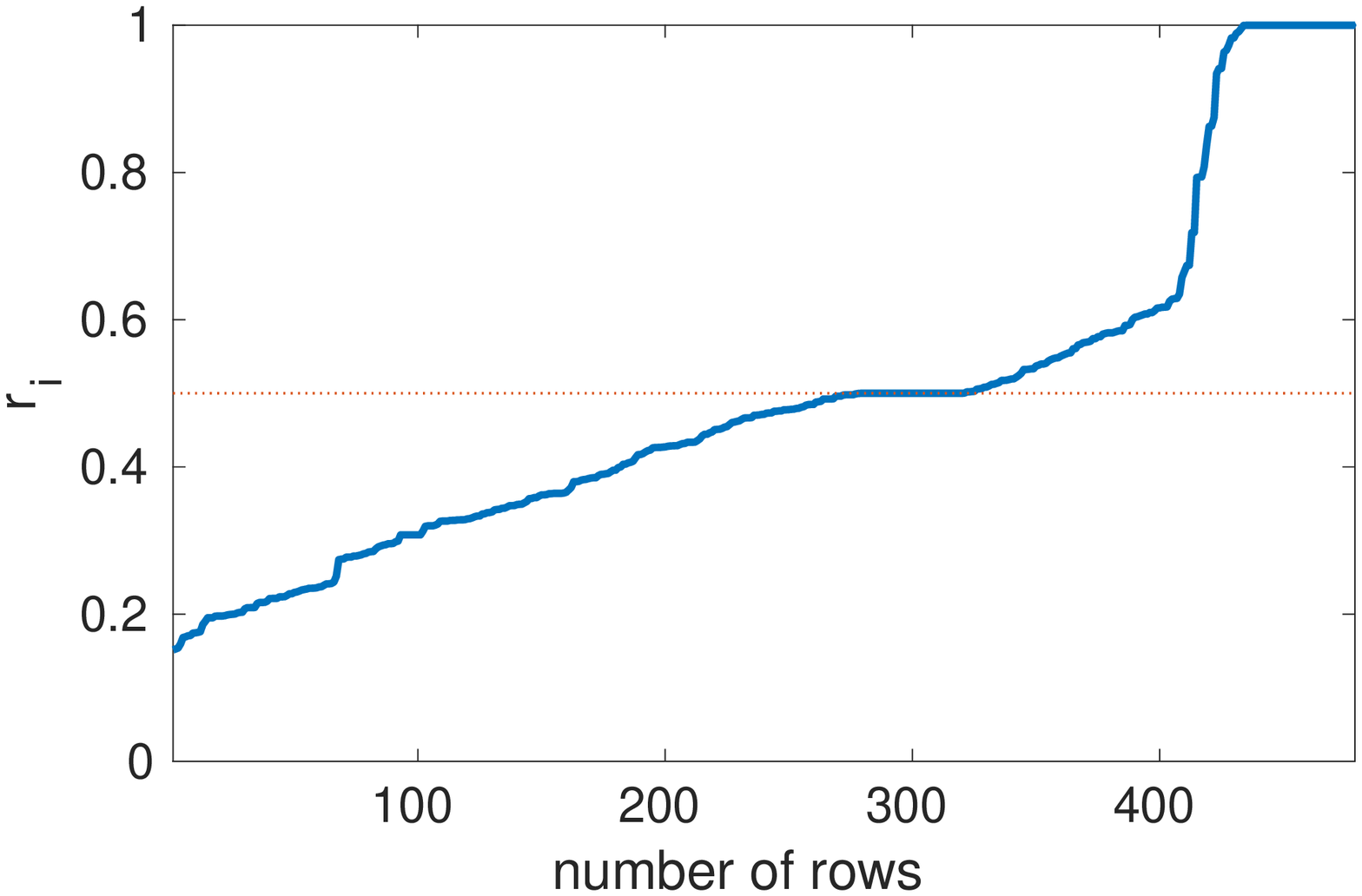} 
 \end{center}  
\end{minipage}
    \caption{Diagonal dominance. Left side: $r_i$ for $A$. Right side: $r_i$  $\tilde A=\Pi^TD_rAD_c$.}
    \label{fig:mwm-dd}
\end{figure}

\section{Symbolic symmetric reordering techniques} 
\label{sec:reordering}
When dealing with large sparse matrices a crucial factor that determines
the computation time is the amount of fill tat is produced during the
factorization of the underlying matrix. To reduce the complexity there
exist many mainly symmetric reordering techniques that attempt to reduce
the fill--in heuristically. Here we will demonstrate only one of these
methods, the so--called nested dissection method. The main reason for selecting
this method is that it can be easily used for parallel computations.

\subsection{Multilevel nested dissection}
\label{subsec:mnd}
Recursive multilevel nested dissection methods for direct
decomposition methods were firstly introduced in the context of
multiprocessing. If parallel direct methods are used to solve a sparse
systems of equations, then a graph partitioning algorithm can be used
to compute a fill reducing ordering that leads to a high degree of
concurrency in the factorization phase. 
\begin{definition}
For a matrix $A\in\R^{n,n}$ we
 define the associated (directed) graph $G_d(A)=(V,E)$, where 
$V=\{1,\dots,n\}$ and the set of edges 
$E=\left\{(i,j)|\, a_{ij}\not=0\right\}$.
The (undirected) graph is given by  $G_d(|A|+|A|^T)$ and is denoted
simply by $G(A)$.
\end{definition}
In graph terminology for a sparse matrix $A$ we simply have a directed
edge $(i,j)$ for any nonzero entry $a_{ij}$ in $G_d(A)$ whereas the 
orientation of the edge is ignored in $G(A)$.

The research on graph-partitioning
methods in the mid-nineteens has resulted in high-quality software
packages, e.g. \metis {} \cite{karypis:98}.
These methods often compute orderings that on the one hand lead small fill--in 
for (incomplete) factorization methods while on the other hand they
provide a high level of concurrency.
We will briefly review the main idea of multilevel nested dissection in
terms graph-partitioning.
\begin{definition}
Let $A\in\R^{n,n}$
and consider its graph $G(A)=(V,E)$. 
A \emph{$k$-way graph partitioning} consists of 
partitionining $V$ into $k$ disjoint subsets
$V_1, V_2, \ldots, V_k$ such that $V_i \cap V_j = \emptyset$ for $i
\ne j$  $\cup_i V_i=V$.
The subset $E_s = E\cap \bigcup_{i\not=j} (V_i\times V_j)$ is called 
\emph{edge separator}.
\end{definition}
Typically we want a $k$-way partitioning to be balanced, i.e., 
each $V_i$ should satisfy $|V_i|\approx n/k$. The edge separator $E_s$
refers to the edges that have to be taken away from the graph
in order to have $k$ separate
subgraphs associated with $V_1,\dots,V_k$ and the number of elements of
$E_s$ is usually referred to as edge-cut. 

\begin{definition}
Given $A\in\R^{n,n}$,
    a \emph{vertex separator} $V_s$ of $G(A)= (V,E)$ is a
    set of vertices such that there exists a $k$-way partitioning 
    $V_1, V_2, \ldots, V_k$ of $V \setminus V_s$ having no edge
    $e\in V_i\times V_j$ for $i\ne j$. 
\end{definition}
A useful vertex separator $V_s$ should not only separate $G(A)$ into
$k$ independent subgraps associated with $V_1,\dots,V_k$, it is 
intended that the numbers of edges 
$\cup_{i=1}^{k} |\{ e_{is} \in V_i, s \in V_s\}| $ is also small.

Nested dissection recursively splits a graph $G(A)= (V,E)$ into almost
equal parts by constructing a vertex separator $V_s$ 
until the desired number $k$ 
of partitionings are obtained. If $k$ is a power of $2$, then a natural
way of obtaining a vertex separator
is to first obtain a $2$-way partitioning of the graph, a so called
\emph{graph bisection} with its associated edge separator $E_s$.
After that a vertex separator $V_s$ is computed from $E_s$, which
gives a $2$-way partitioning $V_1,V_2$ of $V\setminus V_s$.
This process is then repeated separately
for the subgraphs associated with $V_1,V_2$ until eventually a
$k=2^l$-way partitioning is obtained. For the reordering of the
underlying matrix $A$, the vertices associated with $V_1$ are taken first
followed by $V_2$ and $V_s$. This reordering is repeated similarly during
repeated bisection of each $V_i$. In general, vertex separators
of small size result in low fill-in.

\begin{figure}
\sidecaption
{
\begin{minipage}{7.0cm}
{
\begin{minipage}{.6\textwidth}
\includegraphics[width=\textwidth]{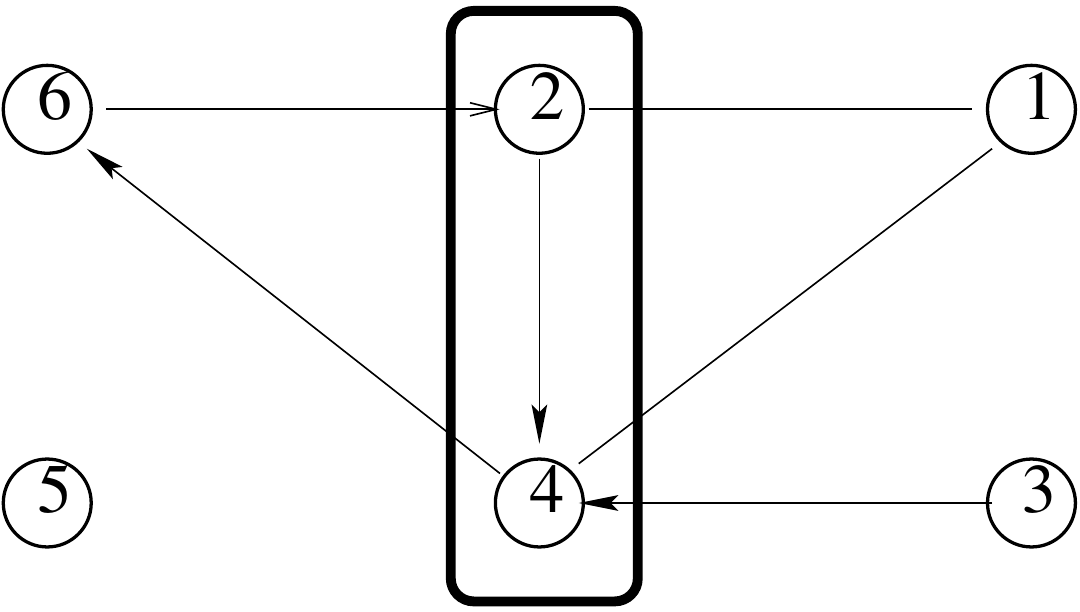}
\end{minipage}
}
\hfil
{
\begin{minipage}{.3\textwidth}
    $\left(
        \begin{array}{cc|cc|cc}
        2   & \nl & \nl & \nl & 4   &  1  \\
        \nl & 3   &  \nl& \nl & \nl & 1   \\ \hline
        \nl &  \nl& 3   & \nl & \nl & \nl \\
        \nl & \nl & \nl &  2  & 1   & \nl \\ \hline
         1  & \nl & \nl & \nl & 3   & 2   \\
        3   & \nl & \nl & 1   & \nl & 4   
        \end{array}
    \right)$
\end{minipage}
}
\end{minipage}
}
\caption{A $2$-way partition with vertex separator $V_s=\{2,4\}$
and the associated reordered matrix placing the two rows and columns associated 
with $V_s$ to the end.}\label{fig:matrixvertex}
\end{figure}

\begin{example}{Vertex Separators}\label{exm:vsep}
To illustrate vertex separators, we consider the reordered matrix $\Pi^TA$
from Figure \ref{fig:unsym_perm} after a  matching is applied.
In Figure \ref{fig:matrixvertex} we display its graph $G(\Pi^T A)$ ignoring
the orientation of the edges. A 
$2$-way partitioning is obtained with $V_1 = \{1,3\}$, $V_2 = \{5,6\}$ and
a vertex separator $V_s = \{2,4\}$. The associated reordering
refers to taking the rows and the columns of $\Pi^T A$ in the order
$1,3,5,6,2,4$.
\end{example}

Since a naive approach to compute a recursive graph bisection is 
typically computationally expensive,
combinatorial \emph{multilevel graph bisection} has been used to
accelerate the process. The basic structure is simple. The multilevel approach
consists of three phases: at first there is a \emph{coarsening phase}
which compresses the given graph successively
level by level by about half of its size. When the coarsest graph with about
a few hundred vertices is reached, the second phase,  namely the so--called
\emph{bisection} is applied. This is a high quality partitioning algorithm.
After that, during the \emph{uncoarsening phase}, the given
bisection is successively refined as it is prolongated towards the original
graph. 

\subsubsection*{Coarsening Phase}
The initial graph $G_0=(V_0,E_0)=G(A)$ of $A\in\R^{n,n}$ is transformed during
the coarsening phase
into a sequence of graphs $G_1, G_2, \ldots, G_m$ of decreasing size 
such that $|V_0|\gg|V_1|\gg|V_2|\gg\cdots\gg|V_m|$. 
Given the graph $G_i=(V_i,E_i)$, the
next coarser graph $G_{i+1}$ is  obtained from $G_i$ by collapsing adjacent
vertices. This can be done e.g. using a maximal matching $\cM_i$ of $G_i$ (cf. Definitions \ref{def:matching} and \ref{def:maxmatching}).
Using $\cM_i$, the next  coarser graph $G_{i+1}$ is 
constructed from $G_i$ collapsing the vertices 
being matched into multinodes, i.e., the elements of  $\cM_i$ together with the
unmatched vertices of $G_i$ become the new vertices $V_{i+1}$ of $G_{i+1}$. 
The new edges $E_{i+1}$ are the remaining edges from $E_i$ 
connected with the collapsed vertices. 
There are various differences in the construction of maximal matchings
\cite{karypis:98,CheP08}. 
One of the most popular and efficient methods is heavy edge
matching \cite{karypis:98}.

\subsubsection*{Partitioning Phase}
At the coarsest level $m$,
a $2$-way partitioning $V_{m,1}\dot{\cup}V_{m,2}=V_m$ of $G_m=(V_m,E_m)$ is computed,
each of them containing about half of the vertices of $G_m$.
This specific partitioning of $G_m$ can be obtained by using various
algorithms such as spectral bisection \cite{fiedler:75} or
combinatorial methods based on Kernighan-Lin variants
\cite{KerL70,FidM97}. It is demonstrated in \cite{karypis:98} that 
for the coarsest graph, combinatorial
methods typically compute smaller edge-cut separators compared with
spectral bisection methods. However, since
the size of the coarsest graph $G_m$ is small (typically $|V_m|<100)$, this
step is negligible with respect to the total amount of computation time.

\subsubsection*{Uncoarsening Phase}
Suppose that at the coarsest level $m$, an edge separator $E_{m,s}$ 
of $G_m$ associated with the  $2$-way partitioning has been computed 
that has lead to a sufficient edge-cut of $G_m$ with $V_{m,1}$, $V_{m,2}$
of almost equal size.
Then $E_{m,s}$ is prolongated to $G_{m-1}$ by reversing the process of
collapsing matched vertices. This leads to an initial edge separator
$E_{m-1,s}$ for $G_{m-1}$. But since $G_{m-1}$ is finer, $E_{m-1,s}$ is 
sub-optimal and one usually decreases the edge-cut of the partitioning
by local refinement heuristics such as the
Kernighan--Lin partitioning algorithm \cite{KerL70} 
or the Fiduccia--Mattheyses method \cite{FidM97}.
Repeating this refinement procedure level--by-level we obtain a sequence
of edge separators $E_{m,s},E_{m-1,s},\dots,E_{0,s}$ and eventually and
edge separator $E_{s}=E_{0,s}$ of the initial graph $G(A)$ is obtained.
If one is seeking for a vertex separator $V_s$ of $G(A)$, then one usually 
computes $V_s$ from $E_s$ at the end.

There have been a number of methods that are used for graph partitioning,
e.g. \metis{} \cite{karypis:98}, a parallel MPI version \parmetis{} \cite{KarSK99},
or a recent multithreaded approach \mtmetis \cite{LasK13}.
Another example for a parallel partitioning algorithm is \scotch \cite{CheP08}. 

\begin{figure}
\sidecaption
\begin{minipage}{.55\textwidth}
  \begin{center}
\includegraphics[width=0.95\textwidth]{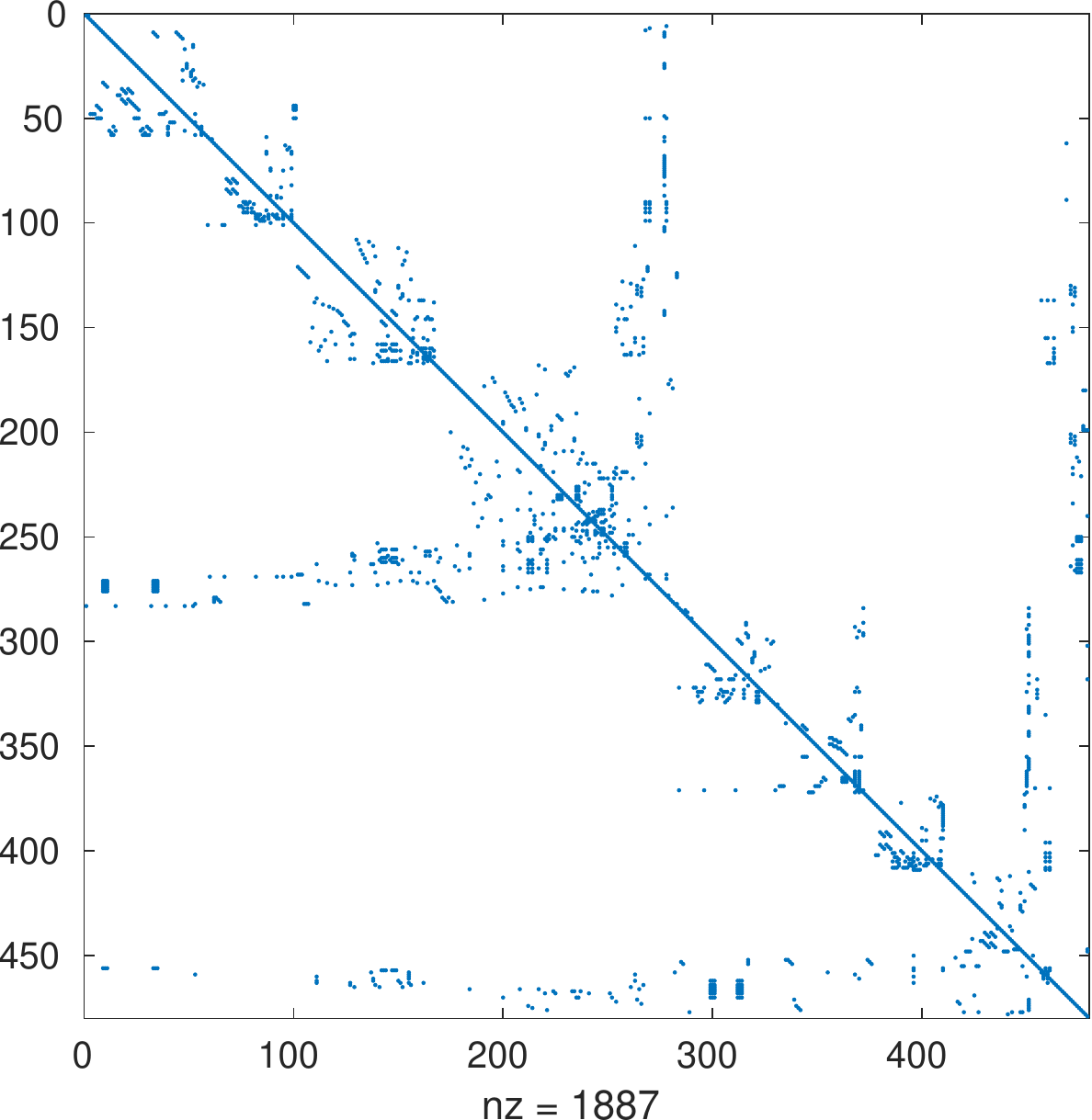} 
 \end{center}  
\end{minipage}
    \caption{Application of multilevel
nested dissection after the matrix is already rescaled and permuted using maximum weight matching.}
    \label{fig:metis}
\end{figure}

\begin{example}{Multilevel Nested Dissection}\label{exm:west0479-metis}
We will continue Example \ref{exm:west0479-match} using the matrix
$\tilde A=\Pi^TD_rAD_s$ that has been rescaled and permuted using
maximum weight matching. We illustrate in Figure \ref{fig:metis} 
how multilevel nested dissection changes the pattern $\hat A=P^T \tilde A P$,
where $P$ refers to the permutation matrix associated with the partitioning
of $G(\tilde A)$.
\end{example}

\subsection{Other reordering methods}
One of the first methods to reorder the system was the
reverse Cuthill--McKee (\rcm)  methods \cite{cm:69,LiuS76} which attempts
to reduce the bandwidth of a given matrix. Though this algorithm is still 
attractive for sequential methods and incomplete factorization methods, its use
for direct solvers is considered as obsolete. An attractive alternative to 
nested dissection as reordering method for direct factorization methods is
the minimum degree algorithm (\mmd) \cite{Ros72,GeoL89} and its recent variants,
in particular the approximate minimum degree algorithm (\amd) \cite{AmeDD96,Dav06} 
with or without constraints. The main objective of the minimum degree algorithm
is to simulate the Gaussian elimination process symbolically by investigating
the update process $a_{ij}\to a_{ij}-a_{ik}a_{kk}^{-1}a_{kj}$ by means of graph
theory, at least in the case of the undirected graph. 
The name-giving degree refers to the number of edges connected to a vertex and 
how the graph and therefore the degrees of its vertices change during the 
factorization process.
Over the years this
has lead to an evolution of the underlying minimum degree algorithm
using the so-called \emph{external degree} for selecting vertices as pivots
and  
further techniques like \emph{incomplete degree update}, 
\emph{element absorption} and \emph{multiple elimination} 
as well as data structures based on cliques. 
For an overview see \cite{GeoL89}.
One of the most costly parts in the minimum degree algorithm  is to update
of the degrees. Instead of computing the exact external degree, in
the approximate minimum degree algorithm
\cite{AmeDD96} an approximate external degree is computed that significantly
saves time while producing comparable fill in the $LU$ decomposition.

We like to conclude this section mentioning that if nested dissection is computed
to produce a vertex separator $V_s$ and a related $k$-way partitioning $V_1,\dots,V-k$ for the remaining vertices of $V\setminus V_s$ of $G(A)=(V,E)$ 
which allow for parallel
computations, then the entries of each $V_i$, $i,\dots,k$ could be taken in
any order. Certainly, inside $V_i$ one could use nested dissection as well, which
is the default choice in multilevel nested dissection methods. However, ass soon
as the coarsest graph $G_m$ is small enough (typically about $100$ vertices),
not only the separator is computed, but in addition the remaining entries of
$G_m$ are reordered to lead to a fill-reducing ordering. In both cases, for
$G_m$ as well as $V_1,\dots,V_k$ one could alternatively use different reordering
methods such as variants of the minimum degree algorithm. Indeed, for
$G_m$ this is what the \metis software is doing. Furthermore, a reordering
method such as the constrained approximate minimum degree algorithm is also
suitable as local reordering for $V_1,\dots,V_k$ as alternative to nested 
dissection, taking into account the edges connected with $V_s$ (also referred to
as HALO structure), see e.g. \cite{PelRA00}.

\section{Sparse $LU$ Decomposition}
\label{sec:lu}
In this section we will assume that the given matrix $A\in\R^{n,n}$ is nonsingular and that it can be factorized as $A=LU$, where $L$ is a lower triangular matrix with unit diagonal and
$U$ is an upper triangular matrix. 
It is well--known \cite{GeoL81}, if $A=LU$, where $L$ and $U^\top$ are lower
triangular matrices, then in the generic case we will have
$G_d(L+U)\supset G_d(A)$, i.e., we will only get additional edges unless some
entries cancel by ``accident'' during the elimination. In the sequel
we will ignore cancellations. Throughout this section we will always assume
that the diagonal entries of $A$ are nonzero as well. We also assume that $G_d(A)$
is connected.

In the preceding sections we have argued that
maximum weight matching often leads to a rescaled and reordered matrix such that
static pivoting is likely to be enough, i.e., 
pivoting is restricted to some dense blocks inside the $LU$ factorization.
Furthermore, reordering strategies such as multilevel nested dissection have 
further symmetrically permuted the system such that the fill--in that occurs
during Gaussian elimination is acceptable and even parallel approaches could
be drawn from this reordering. Thus assuming that $A$ does not need further
reordering and a factorization $A=LU$ exists 
is a realistic scenario in what follows.


\subsection{The Elimination Tree}
\label{subsec:etree}
The basis of determining the fill-in in the triangular factors 
$L$ and $U$ as by-product of the Gaussian elimination can be characterized
as follows (see \cite{Gil94} in the references therein). 

\begin{theorem}\label{thr:gilbert}
Given $A=LU$ with the aforementioned assumptions, 
there exists an edge $(i,j)$ in $G_d(L+U)$ if and only if there exists a path
\[
ix_1, x_2x_3, \dots, x_kj
\]
in $G_d(A)$ such that $x_1,\dots,x_k<\min(i,j)$.
\end{theorem}
In other words, during Gaussian elimination we obtain a fill edge $(i,j)$ for
every path from $i$ to $j$ through vertices less than $\min(i,j)$.

\begin{example}{Fill--in}\label{exm:fill}
We will use the matrix $\Pi^TA$ from Example \ref{exm:vsep}  and sketch
the fill--in obtained during Gaussian elimination   in Figure \ref{fig:fill}.
\end{example}
\begin{figure}
\sidecaption
\begin{minipage}{.45\textwidth}
    $\left(
        \begin{array}{cc|cc|cc}
        2   & \nl    & \nl & \nl & 4   &  1  \\
        \nl & 3      &  \nl& \nl & \nl & 1   \\ \hline
        \nl &  \nl   & 3   & \nl & \nl & \nl \\
        \nl & \nl    & \nl &  2  & 1   & \nl \\ \hline
         1  & \nl    & \nl & \nl & 3   & 2   \\
         3  & \nl    & \nl & 1   & \times & 4   
        \end{array}
    \right)$
\end{minipage}
    \caption{Fill-in with respect to $L+U$ is denoted by $\times$.}
    \label{fig:fill}
\end{figure}


A problem in general is to predict the filled graph $G_d(L+U)$ and the fastest
known method to compute it, is Gaussian elimination.
The situation simplifies if the graph is undirected. 
In the sequel we ignore the orientation of the edges and simply consider
the undirected graph $G(A)$ and $G(L+U)$, respectively.

\begin{definition}
The undirected graph $G(L+U)$ that is derived from the undirected graph 
$G(A)$ by applying Theorem \ref{thr:gilbert} is called the \emph{filled graph} 
and it will be denoted by $G_f(A)$.
\end{definition}

\begin{example}{Fill-in with respect to the undirected graph}\label{exm:symfill}
When we consider the undirected graph $G(A)$ in Example \ref{exm:fill},
the pattern of $|\Pi^TA|+|\Pi^TA|^T$ and its filled graph $G_f(A)$ now equals 
$G(A)$
(cf. Figure \ref{fig:symfill}).
\end{example}
\begin{figure}
\sidecaption
\begin{minipage}{.45\textwidth}
    $\left(
        \begin{array}{cc|cc|cc}
         \bullet&\nl    & \nl   & \nl   &\bullet&  \bullet  \\
         \nl    &\bullet&  \nl  & \nl   & \nl   & \bullet   \\ \hline
         \nl    &\nl    &\bullet& \nl   &\nl    & \nl \\
         \nl    &\nl    & \nl   &\bullet&\bullet& \bullet \\ \hline
         \bullet&\nl    & \nl   &\bullet&\bullet& \bullet   \\
         \bullet&\bullet& \nl   &\bullet&\bullet& \bullet   
        \end{array}
    \right)$
\end{minipage}
    \caption{Entries of $G(A)$ are denoted by $\bullet$.}
    \label{fig:symfill}
\end{figure}

The key tool to predict the fill--in easily for the undirected graph is the
\emph{elimination tree} \cite{Liu90}. 

Recall that an undirected and connected
graph is called a \emph{tree}, if it does not contain any cycle.
Furthermore, one vertex is identified as \emph{root}.
As usual we call a vertex $j$ \emph{parent} of $i$, if there exists an edge
$(i,j)$ in the tree such that $j$ is closer to the root. In this case
$i$ is called \emph{child} of $j$. The subtree rooted at vertex $j$ is denoted
by $T(j)$ and the vertices of this subtree 
are called \emph{descendants} of $j$ whereas $j$ is called their \emph{ancestor}.
Initially we will define the elimination tree algorithmically 
using the depth--first--search algorithm \cite{AhoHU83}. Later we will
state a much simplified algorithm. 
\begin{definition}\label{def:etree}
Given the filled graph $G_f(A)$ the
\emph{elimination tree} $T(A)$ is defined by the following algorithm.\\
Perform a depth--first--search in $G_f(A)$ starting
from vertex $n$.\\  
When vertex $m$ is visited, choose
from its unvisited neighbors $i_1,\dots,i_k$ the index
$j$ with the largest number $j=\max\{i_1,\dots,i_k\}$ and
continue the search with $j$. \\
A leaf of the tree is reached, 
when all neighbors have already been visited.
\end{definition}
We like to point out that the application of the 
depth--first--search to $G_f(A)$ starting at vertex $n$ behaves
significantly different from other graphs.
By Theorem \ref{thr:gilbert} it follows that
as soon as we visit a vertex $m$, all its neighbors $j>m$ must have been 
visited prior to vertex $m$. Thus  the labels of the vertices are strictly 
decreasing until we reach a leaf node.

\begin{figure}[htb]
{
\begin{minipage}{.35\textwidth}
\includegraphics[width=\textwidth]{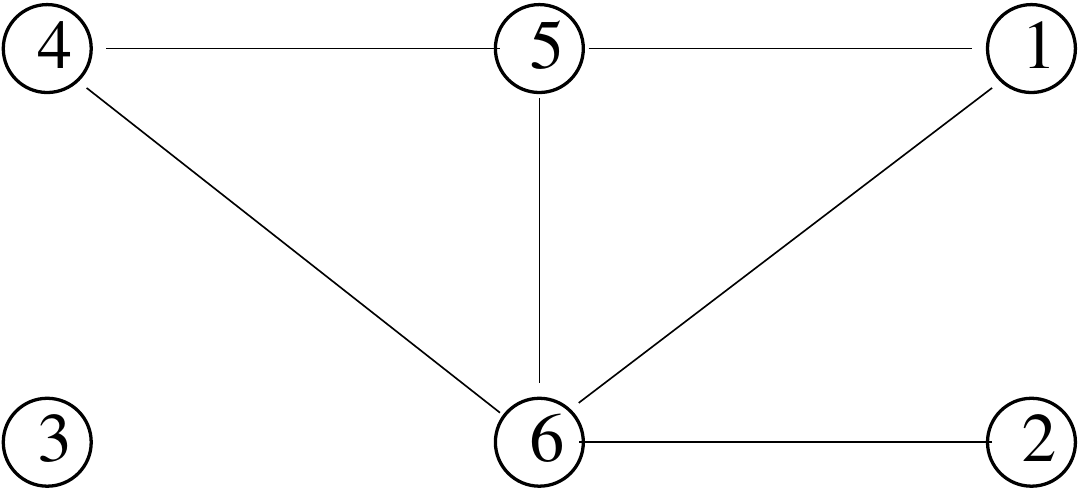}
\end{minipage}
}
~~~~\hfil~~~~
{
\begin{minipage}{.35\textwidth}
\includegraphics[width=\textwidth]{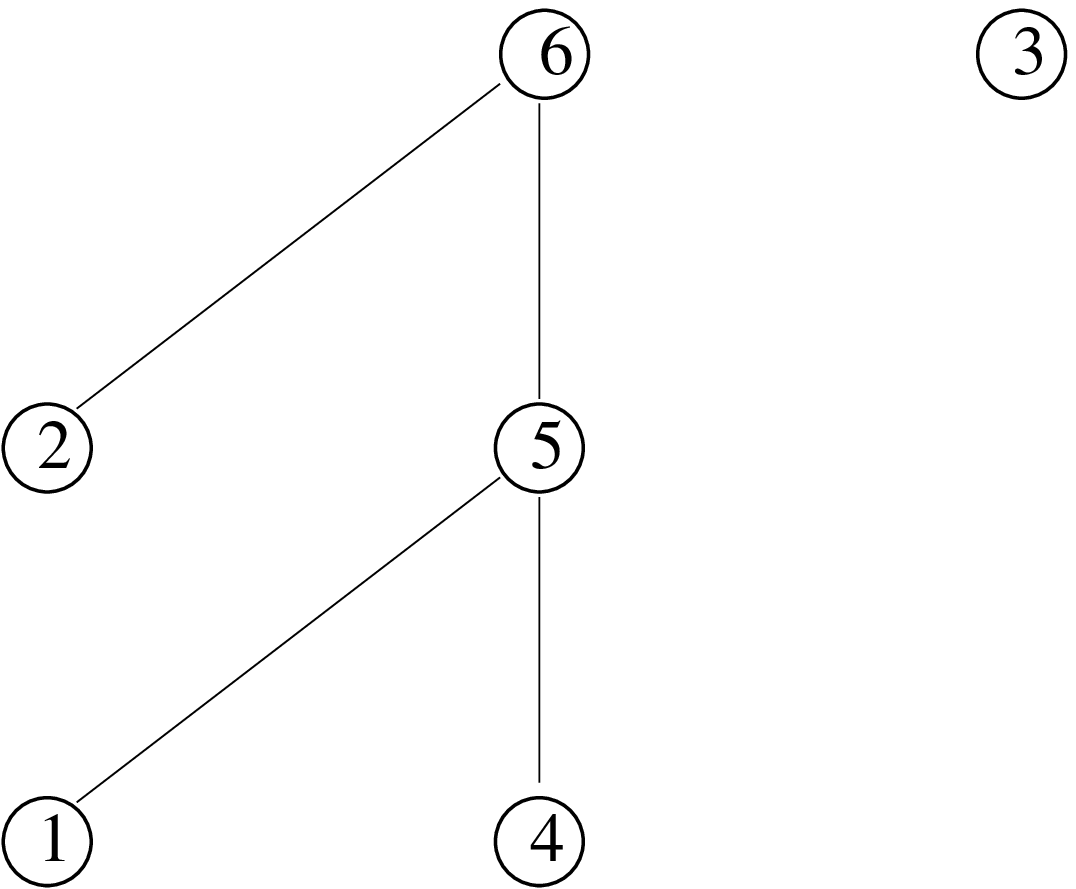}
\end{minipage}
}
\caption{Filled graph (left) and elimination tree(s) (right).}\label{fig:matrixpattern}
\end{figure}

\begin{example}{Depth--first--search}\label{exm:dfs}
We illustrate the depth--first--search using the filled graph in
Figure \ref{fig:matrixpattern} and the pattern from Example \ref{exm:symfill}.
The extra fill edge is marked by the bold line. 

The ongoing depth--first search visits the vertices in the order
$6\to5\to4$. Since at vertex $4$, all neighbors of $4$ are visited (and indeed have a larger number), the algorithm backtracks to $5$ and continues the search in the order
$5\to1$. Again all neighbors of vertex $1$ are visited (and have larger number), 
thus the algorithm backtracks to $5$ and to $6$ and continues by $6\to2$. Then the
algorithm terminates. Note that vertex $3$ is isolated and if the graph of $A$
is not connected one has to proceed for each connected component separately.
\end{example}

\begin{remark}\label{rem:cross-edges}
It follows immediately from the construction of $T(A)$ and Theorem 
\ref{thr:gilbert} 
that additional edges of $G_f(A)$ which are not covered by the elimination
tree can only show up between a vertex and some of its ancestors (referred to as ``back--edges''). In contrast to that, ``cross--edges'' between unrelated vertices
do not exist. 
\end{remark}

\begin{remark}\label{rem:dependence}
One immediate consequence of Remark \ref{rem:cross-edges} is 
that triangular factors can be computed independently starting from the
leaves until the vertices meet a common parent, i.e., 
column $j$ of $L$ and $U^T$ only depend on those columns $s$
of $L$ and $U^T$ such that $s$ is a descendant
of $j$ in the elimination tree $T(A)$.
\end{remark}

\begin{figure}
{
\begin{minipage}{.99\textwidth}
\includegraphics[width=\textwidth,height=0.3\textwidth]{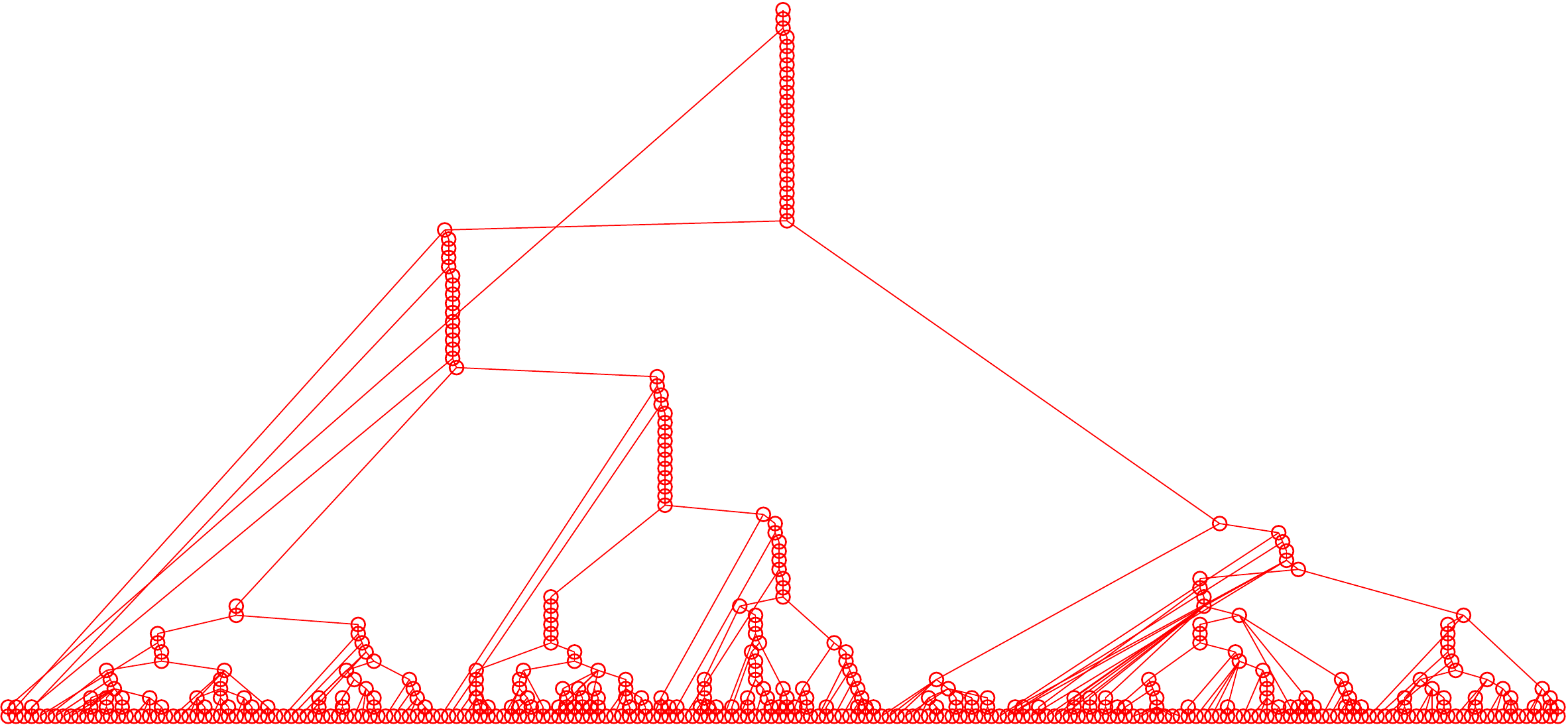}
\end{minipage}
}
\caption{Elimination tree of ``west0479'' after maximum weight matching and nested dissection are applied.}\label{fig:etreeplot}
\end{figure}

\begin{example}{Elimination tree}\label{exm:etree}
We use the matrix ``west0479'' from Example \ref{exm:west0479-metis},
after maximum weight matching and multilevel nested dissection have
been applied. We use \ml's \texttt{etreeplot} to display its elimination
tree (see Figure \ref{fig:etreeplot}). The elimination tree displays the high
level of concurrency that is induced by nested dissection, since 
by Remark \ref{rem:dependence} the computations can be executed independently
at each leaf node towards to the root until a common parent vertex is reached.
\end{example}

Further conclusions can be easily derived
from the elimination tree, in particular
Remark   \ref{rem:dependence} in conjunction with Theorem \ref{thr:gilbert}. 
\begin{remark}\label{rem:path-compression}
Consider some $k\in\{1,\dots,n\}$. Then there exists a (fill) edge $(j,k)$ 
with $j<k$ if and only if there exists a common
descendant $i$ of $k,j$ in $T(A)$ such that $a_{ik}\not=0$.
This follows from the fact that once $a_{ik}\not=0$, by Theorem \ref{thr:gilbert}
this induces (fill) edges $(j,k)$ in the filled graph $G_f(A)$ for all nodes
$j$ between $i$ and $k$ in the elimination tree $T(A)$, i.e., for all ancestors
of $i$ that are also descendents of $k$. This way, $i$ propagates fill--edges
along the branch from $i$ to $k$ in $T(A)$ and the information $a_{ik}\not=0$ can be
used as path compression to advance from $i$ towards to $k$ along the elimination
tree.
\end{remark}
\begin{example}{Path compression}\label{exm:pc}
Consider the graph and the elimination tree from 
Figure \ref{fig:matrixpattern}. Since there exists the edge $(1,6)$ in $G(A)$,
therefore another (fill) edge $(5,6)$ must exist (here not a fill edge, but a regular edge). Similarly, the same conclusion can be drawn
from the existence of the edge $(4,6)$.
Via $a_{16}\not=0$ we can advance from vertex $1$ to vertex $6$ bypassing vertex $5$.
\end{example}

The elimination tree itself can be easily described by a vector $p$ of length
$n$ such that for any $i<n$, $p_i$ denotes the parent node while $p_n=0$
corresponds to the root. 
Consider some step $k$ with $a_{ik}\not=0$, for some $i<k$. 
By Remark \ref{rem:path-compression}, $i$ must be a descendent of $k$
and there could be further ancestors $j$ of $i$ which are also descendents of $k$.
Possibly not all ancestors of $i$ have been assigned a parent node so far.
Thus we can replace $i$ by $j=p_i$ until we end up with $p_j=0$ or $p_j\geqslant k$.
This way we traverse $T(A)$ from $i$ towards to $k$ until we have found the child 
node $j$ of $k$. If the parent of $j$ has not been assigned to $j$ yet, 
then $p_j=0$ and $k$ must be the parent of $j$. If some $l<k$ were the parent of 
$j$, then we would have assigned $l$ as parent of $j$ in an earlier step $l<k$.
In this case we set $p_j\leftarrow k$. Otherwise, if $p_j\geqslant k$, then we have already
assigned $j$'s parent in an earlier step $l<k$.

\begin{example}{Computation of parent nodes}
Consider the elimination tree $T(A)$ from Figure \ref{fig:matrixpattern}.
Unless $k=5$, no parents have been assigned, i.e. $p_i=0$ for all $i$.

Now for $k=5$ we have 
$a_{15}\not=0$ and using the fact that $p_1=0$ implies that we have to set
$p_1\leftarrow 5$. Next, $a_{45}\not=0$ and again $p_4=0$ requires to set $p_4\leftarrow 5$.

Finally, if $k=6$, we have $a_{16}\not=0$, we advance from $1$ to $p_1=5$ and
since $p_5=0$ we set $p_5\leftarrow 6$. Next, $a_{26}\not=0$ and $p_2=0$ imply
that we have to set $p_2\leftarrow 6$. $a_{46}\not=0$ will traverse from $4$ to
$p_4=5$ to $p_5=6$, which requires no further setting. Last but not least,
$a_{56}\not=0$ requires no further action, since we already have $p_5=6$.

In total we have $p=[5, 6, 0, 5, 6, 0]$ which perfectly reveals the parent
properties of the elimination trees in  Figure \ref{fig:matrixpattern}.
\end{example}

By Remark \ref{rem:path-compression} (cf. \cite{Tar83,Dav06}), 
we can also make use of path compression.
Since our goal is to traverse the branch of the elimination tree from $i$ to $k$
as fast as possible, any ancestor $j=a_i$ of $i$ would be sufficient. With the
same argument as before, an ancestor $a_j=0$ would refer to a vertex that does
not have a parent yet. In this case we can again set $p_j\leftarrow k$. Moreover,
$k$ is always an ancestor of $a_i$.

The algorithm including path compression can be summarized as follows 
(see also  \cite{Liu90,Dav06}).
\begin{programcode}{Computation of the elimination tree}\label{alg:etree}
\begin{algorithmic}[1]
  \Require $A\in\R^{n,n}$ such that $A$ has the same pattern as $|A|+|A|^T$.
  \Ensure vector $p\in\R^n$ such that $p_i$ is the parent of $i$, $i=1,\dots,n-1$,
except $p_n=0$.
  \State let $a\in\R^n$ be an auxiliary vector used for path compression.
  \State $p\leftarrow 0, a\leftarrow 0$
    \For{$k=2,\dots,n$}
        \For{all $i<k$ such that $a_{ik}\not=0$}
            \While{$i\not=0$ and $i<k$}
                  \State $j\leftarrow a_i$
                  \State $a_i\leftarrow k$
                  \If{$j=0$} 
                     \State $p_i\leftarrow k$
                  \EndIf   
                  \State $i\leftarrow j$
            \EndWhile
        \EndFor
    \EndFor
\end{algorithmic}
\end{programcode}

\subsection{The supernodal approach}
We have already seen that the elimination tree reveals information about
concurrency. It is further useful to determine the fill-- in $L$ and $U^T$.
This information can be computed from the elimination tree $T(A)$ together
with $G(A)$. The basis for determining the fill--in in each column is 
again Remark \ref{rem:path-compression}. Suppose we are interested in the
nonzero entries of column $j$ of $L$ and $U^T$. Then for all decedents of $j$,
i.e. the nodes of the subtree $T(j)$ rooted at vertex $j$, a nonzero entry
$a_{ik}\not=0$ also implies $l_{kj}\not=0$. Thus, starting at any leaf $i$,
we obtain its fill by all $a_{ik}\not=0$ such that $k>i$ and when we move forward
from $i$ to its parent $j$, vertex $j$ will inherit the fill from node $i$ for
all $k>j$ plus the nonzero entries given by $a_{jk}\not=0$ such that $k>j$.
When we reach a common parent node $k$ with multiple children, the same argument
applies using the union of fill--in greater than $k$ from its children together
with the nonzero entries $a_{kl}\not=0$ such that $l>k$.
We summarize this result in a very simple algorithm
\begin{programcode}{Computation of fill--in}\label{alg:compute_pattern}
\begin{algorithmic}[1]
  \Require $A\in\R^{n,n}$ such that $A$ has the same pattern as $|A|+|A|^T$.
  \Ensure sparse strict lower triangular pattern $P\in\R^{n,n}$ with
  same pattern as $L$, $U^T$.
  \State compute parent array $p$ of the elimination tree $T(A)$
  \For{$j=1,\dots,n$}
      \State supplement nonzeros of column $j$ of $P$ with all $i>j$ such that $a_{ij}\not=0$ 
      \State $k=p_j$
      \If{$k>0$} 
         \State supplement nonzeros of column $k$ of $P$ with nonzeros of column $j$ of $P$ greater than $p$
      \EndIf   
  \EndFor
\end{algorithmic}
\end{programcode}

Algorithm \ref{alg:compute_pattern} only deals with the fill pattern. 
One additional aspect that allows to raise efficiency
and to speed up the numerical factorization significantly 
is to detect dense submatrices in the factorization.
Block structures  allow to collect parts
of the matrix in dense blocks and to treat them commonly using 
dense matrix kernels such as level--3 BLAS and LAPACK \cite{DodL85,DonDHH88}.

Dense blocks can be read off from the elimination tree employing
Algorithm \ref{alg:compute_pattern}.
\begin{definition}\label{def:supernode}
Denote by $\mathcal{P}_j$ the nonzero indices of column $j$ of $P$
as computed by Algorithm \ref{alg:compute_pattern}.
A sequence $k,k+1,\dots,k+s-1$ is called \emph{supernode} of size $s$
if the columns of $\mathcal{P}_{j}=\mathcal{P}_{j+1}\cup \{j+1\}$
for all $j=k,\dots,k+s-2$.
\end{definition}
In simple words, Definition \ref{def:supernode} states that for a supernode
$s$ subsequent columns can be grouped together in one dense block with a triangular
diagonal block and a dense subdiagonal block since they perfectly match the 
associated trapezoidal shape. We can thus easily supplement 
Algorithm \ref{alg:compute_pattern} with a supernode detection.
\begin{programcode}{Computation of fill--in and supernodes}\label{alg:compute_supernode}
\begin{algorithmic}[1]
  \Require $A\in\R^{n,n}$ such that $A$ has the same pattern as $|A|+|A|^T$.
  \Ensure sparse strict lower triangular pattern $P\in\R^{n,n}$ with
  same pattern as $L$, $U^T$ as well as column size $s\in\R^m$ of each supernode.
  \State compute parent array $p$ of the elimination tree $T(A)$
  \State $m\leftarrow0$ 
  \For{$j=1,\dots,n$}
      \State supplement nonzeros of column $j$ of $P$ with all $i>j$ such that $a_{ij}\not=0$ 
      \State denote by $r$ the number of entries in column $j$ of $P$
      \If{$j>1$ and $j=p_{j-1}$ and $s_m+r=l$}
         \State $s_m\leftarrow s_m+1$ \Comment{continue current supernode}
      \Else
         \State $m\leftarrow m+1$, $s_m\leftarrow 1$, $l\leftarrow r$ \Comment{start new supernode}
      \EndIf
      \State $k=p_j$
      \If{$k>0$} 
         \State supplement nonzeros of column $k$ of $P$ with nonzeros of column $j$ of $P$ greater than $p$
      \EndIf   
  \EndFor
\end{algorithmic}
\end{programcode}

\begin{figure}
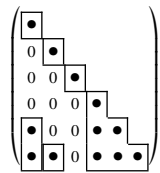

\sidecaption
\begin{minipage}{.45\textwidth}
    $\left(
        \begin{array}{cccccc}
         \\[-1.5ex]\cline{1-1}
         \multicolumn{1}{|c|}{\bullet}&       &       &       &       &         \\ \cline{1-2}
         \nl    &\multicolumn{1}{|c|}{\bullet}&       &       &       &         \\ \cline{2-3}
         \nl    &\nl    &\multicolumn{1}{|c|}{\bullet}&       &       &         \\ \cline{3-4}
         \nl    &\nl    & \nl   &\multicolumn{1}{|c|}{\bullet}&       &         \\ \cline{1-1}\cline{5-5}
         \multicolumn{1}{|c|}{\bullet}&\nl    & \nl   &\multicolumn{1}{|c}{\bullet}&\multicolumn{1}{c|}{\bullet}&         \\ \cline{2-2}\cline{6-6}
         \multicolumn{1}{|c|}{\bullet}&\multicolumn{1}{|c|}{\bullet}& \nl   &\multicolumn{1}{|c}{\bullet}&\bullet& \multicolumn{1}{c|}{\bullet}\\
\cline{1-1}\cline{2-2}\cline{4-6}
 \end{array}
    \right)$
\end{minipage}
    \caption{Supernodes in the triangular factor.}
    \label{fig:supernode}
\end{figure}

\begin{example}{Supernode Computation}
To illustrate the use of supernodes, we consider the matrix pattern
from Figure \ref{fig:symfill} and illustrate the underlying
dense block structure in Figure \ref{fig:supernode}.
Supernodes are the columns $1$, $2$, $3$ as scalar columns as well as columns
$4$--$6$ as one single supernode.
\end{example}

\begin{figure}
\begin{minipage}{.48\textwidth}
 \begin{center}
\includegraphics[width=0.99\textwidth]{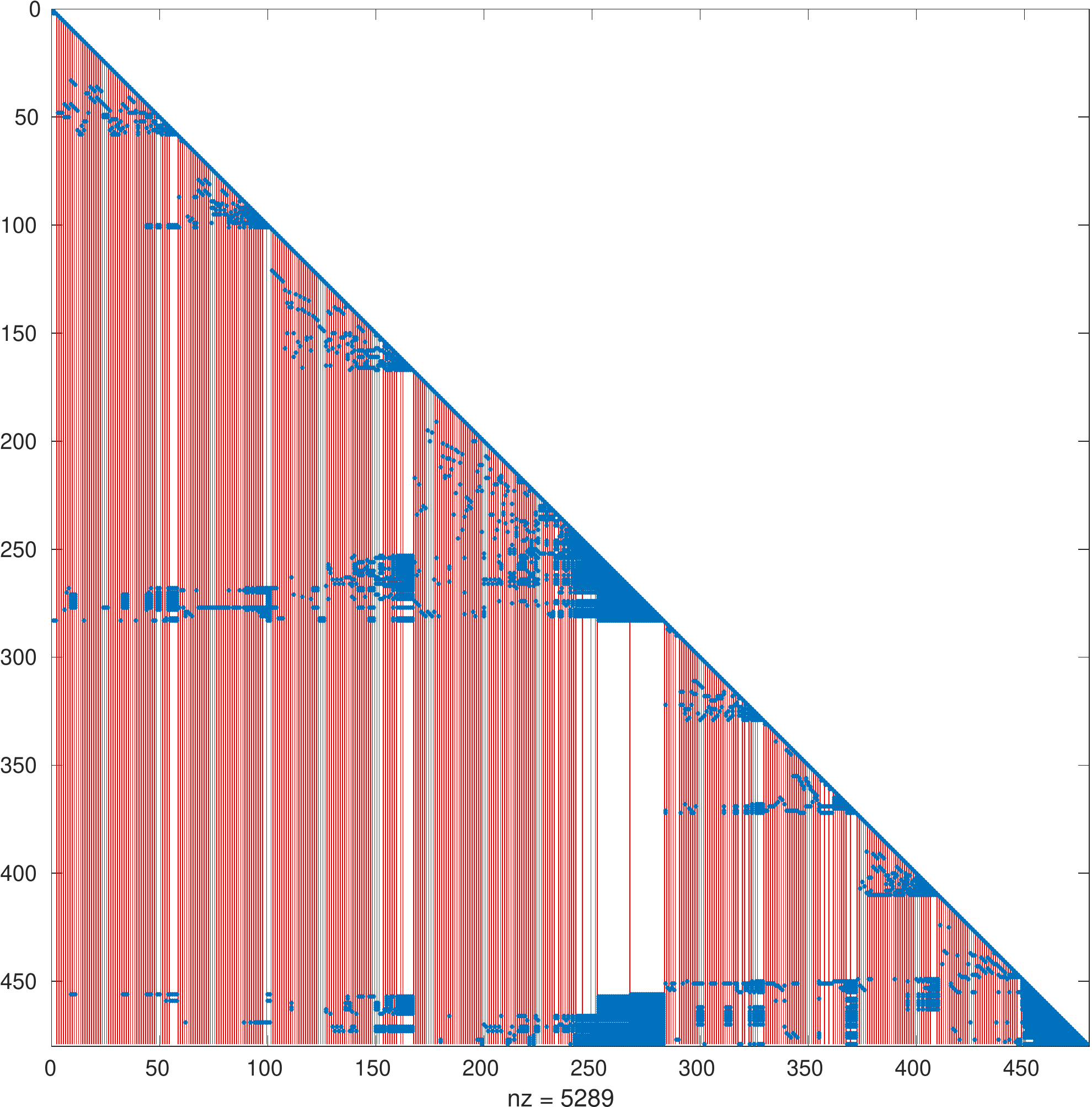} 
 \end{center}
\end{minipage}
~
\begin{minipage}{.48\textwidth}
  \begin{center}
\includegraphics[width=0.99\textwidth]{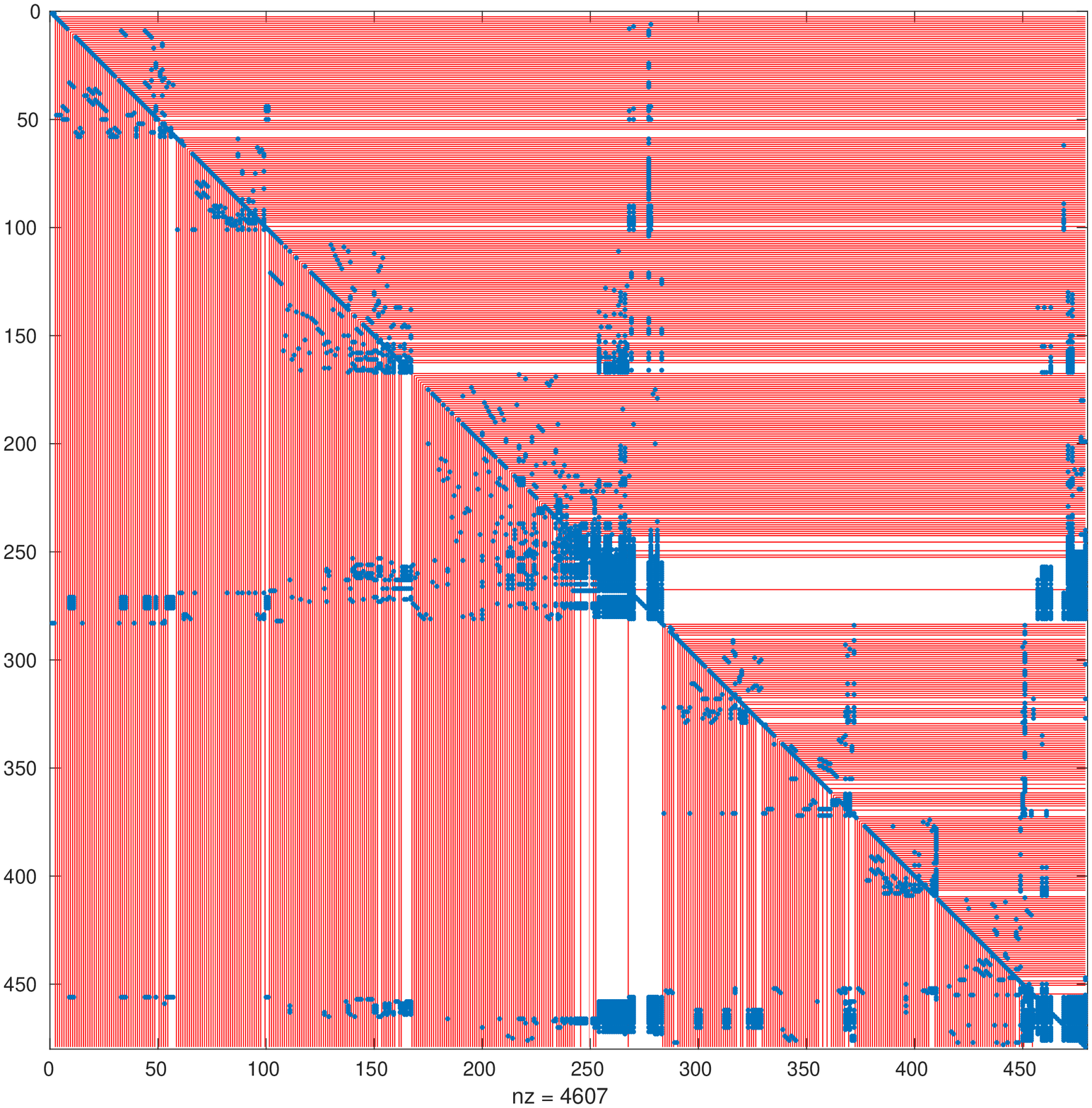} 
 \end{center}  
\end{minipage}
    \caption{Supernodal structure. Left: vertical lines display the blocking of the
supernodes with respect to the associated Cholesky factor. Right: 
vertical and horizontal lines display the blocking of the
supernodes applied to $L$ and $U$.}
    \label{fig:supernodal_structure}
\end{figure}

Supernodes form the basis of several improvements, e.g.,
a supernode can be stored as one or two dense matrices. 
Beside the storage scheme as dense matrices, the nonzero row indices
for these blocks need only be stored once.
Next the use of dense submatrices allows the usage of dense matrix kernels
using level--3 BLAS
\cite{DodL85,DonDHH88}. 
\begin{example}{Supernodes}\label{exm:supernodes}
We use the matrix ``west0479'' from Example \ref{exm:west0479-metis},
after maximum weight matching and multilevel nested dissection have
been applied. 
We use its undirected graph to compute the supernodal
structure. Certainly, since the matrix is nonsymmetric, the block structure
is only sub-optimal. We display the supernodal structure for the associated
Cholesky factor, i.e., for the Cholesky factor of an symmetric positive definite
matrix with same undirected graph as our matrix (see
left part of Figure \ref{fig:supernodal_structure}). Furthermore, we display
the supernodal structure for the factors $L$ and $U$ computed from the
nonsymmetric matrix without pivoting (see right part of Figure \ref{fig:supernodal_structure}).
\end{example}

While the construction of supernodes is fairly easy in the symmetric case,
its generalization to the general case is significantly harder, since one
has to deal with pivoting in each step of Gaussian elimination.
In this case one uses the column elimination tree \cite{GeoN85}.


\section{Sparse Direct Solvers --- Supernodal Data Structures}
\label{sec:parallel}

\begin{figure*}[t]
    \centering
    \begin{minipage}{.35\textwidth}
        \centering
        \includegraphics[width=0.85\textwidth,clip=true]{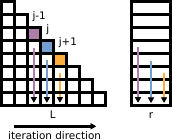}
    \end{minipage}%
    \begin{minipage}{0.65\textwidth}
        \centering
  \begin{algorithmic}[1]
    \Procedure{Sparse forward substitution}{}
            \For{j = 0; j < n; j++}\label{algo:fw:cholmod}
                \For{i = \nxlnz[j]; i < \nxlnz[j+1]; i++}
                   \State row = \nindx[i]
                   \State \nr[row] -=  \nr[j] * \nlnz[i] \Comment{indexed DAXPY}
            \EndFor\label{algo:fw:cholmod:rloop:end}
      \EndFor
    \EndProcedure
  \end{algorithmic}
    \end{minipage}

\bigskip

    \centering
    \begin{minipage}{.35\textwidth}
        \centering
        \includegraphics[width=0.85\textwidth,clip=true]{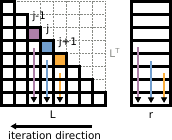}
    \end{minipage}%
    \begin{minipage}{0.65\textwidth}
        \centering
  \begin{algorithmic}[1]
    \Procedure{Sparse backward substitution}{}
            \For{j =  n;  j > 0; j - -}\label{algo:bw:cholmod}
                \For{i = \nxlnz[j]; i < \nxlnz[j+1]; i++}
                   \State row = \nindx[i]
                   \State \nr[j] -=  \nr[row] * \nlnz[i] \Comment{indexed DDOT}
            \EndFor\label{algo:bw:cholmod:rloop:end}
      \EndFor
    \EndProcedure
  \end{algorithmic}
    \end{minipage}
  \caption{Sparse triangular substitution in CSC format based on indexed DAXPY/DDOT kernel operations.}
  \label{algo:triangular}
\end{figure*}

\begin{figure}[t]
  \centering
    \includegraphics[width=0.5\textwidth,clip=true]{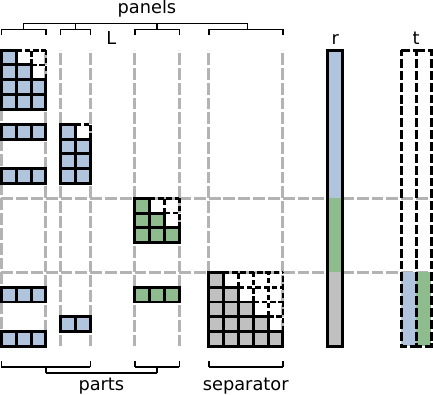}
  \caption{Sparse matrix data structures in PARDISO. Adjacent columns of $L$ exhibiting the same
structure form panels also known as supernodes. 
Groups of panels which touch independent elements of the right hand side $r$ are
parts. The last part in the lower triangular matrix $L$ is called separator.}
  \label{fig:algo:ds}
\end{figure}

High-performance sparse solver libraries have been a very important part of
scientific and engineering computing for years, and their importance
continues to grow as microprocessor architectures become more complex
and software libraries become better designed to integrate easily
within applications. Despite the fact that there are various science
and engineering applications, the underlying algorithms typically have
remarkable similarities, especially those algorithms that are most
challenging to implement well in parallel. It is not too strong a
statement to say that these software libraries are essential to the
broad success of scalable high-performance computing in computational
sciences.  In this section we demonstrate the benefit of supernodal data structures within the 
sparse solver package PARDISO~\cite{schenk-2004}. We illustrate it by using 
the triangular solution process. The forward and backward substitution is performed
column wise with respect to the columns of $L$, starting with the
first column, as depicted in Figure~\ref{algo:triangular}.
The data dependencies here allow to store vectors $y$, $z$, $b$, and $x$ in only one
vector $r$. When column $j$ is reached, $r_j$ contains the solution for $y_j$. 
All other elements of $L$ in this column, i.\,e.\ $L_{ij}$ with $i = j + 1,
\ldots, N$, are used to update the remaining entries in $r$ by 
\be
  r_i = r_i - r_j L_{ij}.
  \label{eq:algo:fw:pardiso}
\ee
The backward substitution with~$L^T$ will take place row wise, since we
use $L$ and perform the substitution column wise with respect to $L$, as shown in the lower part of
Figure~\ref{algo:triangular}.  In contrast to the forward substitution the
iteration over columns starts at the last column $N$ and proceeds to
the first one.  If column $j$ is reached, then $r_j$, which contains the $j$-component of the solution vector $x_j$,
is computed by subtracting the dot-product of the remaining elements in
the column $L_{ij}$ and the corresponding elements of $r_i$ with $i =
j + 1, \ldots, N$ from it:
\be
  r_j = r_j - r_i L_{ij} .
  \label{eq:algo:bw:pardiso}
\ee
After all columns have been processed $r$ contains the required solution of $x$. It is important to note that
line 5 represents in both substitutions an indexed DAXPY and indexed
DDOT kernel operations that has to be computed during the streaming 
operations of the vector $r$ and the column $j$ of the numerical factor $L$. 
As we are dealing with sparse matrices it makes no sense to store the lower
triangular matrix $L$ as a dense matrix.
Hence PARDISO uses its own data structure to store $L$, as shown in
Figure~\ref{fig:algo:ds}. 

\begin{algorithm}[t]
  \begin{algorithmic}[1]
    \Procedure{Forward}{}
      \For{part $o$ in parts} \Comment{parallel execution}
        \For{panel p in part $p$}
          \For{\textcolor{blue}{column $j$ in panel}} \Comment{unroll} \label{alg:fw:1}
            \State i = \nxindx{}[p] + offset
          
            \For{k = \nxlnz[j] + offset; k < sep; ++k}\label{algo:fw:rloop}
                \State row = \nindx[i++]
                \State \nr[row] - =  \nr[j] \nlnz[k] \Comment{indexed DAXPY} 
            \EndFor\label{algo:fw:rloop:end}
            \For{k = sep + 1; k < \nxlnz[j+1]; ++k}\label{algo:fw:seploop}
                \State row = \nindx[i++]
                \State \ntemp[row,p] -=  \nr[j] \nlnz[k] \Comment{indexed DAXPY}
            \EndFor\label{algo:fw:seploop:end}
          \EndFor
        \EndFor
      \EndFor
      \State r[i] = r[i] - sum(\ntemp[i,:])  \Comment{gather temporary arrays}
      \For{panel p in separator} \Comment{serial execution}
        \For{\textcolor{blue}{column $j$ in panel}} \Comment{unroll}\label{alg:fw:2}
            \State i = \nxindx[p] + offset
          
            \For{k = \nxlnz[j] + offset; k < \nxlnz[j+1]; ++k}
                \State row = \nindx[i++]
                \State \nr[row] -=  \nr[j] \nlnz[k] \Comment{indexed DAXPY}
            \EndFor
        \EndFor
      \EndFor
    \EndProcedure
  \end{algorithmic}
  \caption{Forward substitution in PARDISO. Note that in case of serial
execution separated updates to temporary arrays in line
\ref{algo:fw:seploop}--\ref{algo:fw:seploop:end} are not necessary
and can be handled via the loop in lines
\ref{algo:fw:rloop}--\ref{algo:fw:rloop:end}.}
  \label{alg:algo:fw}
\end{algorithm}

Adjacent columns exhibiting the same row sparsity structure form a \textit{panel}, also known
as \textit{supernode}.
A panel's column count is called the \textit{panel size} $n_p$.
The columns of a panel are stored consecutively in memory excluding the zero
entries. 
Note that columns of panels are padded in the front with zeros so they get the 
same length as the first column inside their panel. The padding is of utmost performance
for the PARDISO solver to use Level-3 BLAS and LAPACK functionalities~\cite{20.500.11850/144477}.
 Furthermore panels are stored consecutively in the \vlnz{} array. 
Row and column information is now stored in accompanying arrays.
The \texttt{xsuper} array stores for each panel the index of its first column. 
Also note that here column indices are the running count of nonzero columns.
Column indices are used as indices into \vxlnz{} array to lookup the start of
the column in the \vlnz{} array which contains the numerical values of the factor $L$.
To determine the row index of a column's element an additional array \vindx{} is
used, which holds for each panel the row indices.
The start of a panel inside \vindx{} is found via \vxindx{} array.
The first row index of panel~$p$ is \vindx\texttt{[\vxindx[p]]}.
For serial execution this information is enough. 
However, during parallel forward/backward substitution concurrent updates to
the same entry of \vr{} must be avoided.
The \textit{parts} structure contains the start (and end) indices of the panels which can
be updated independently as they do not touch the same entries of $r$.
Two parts, colored blue and orange, are shown in Figure~\ref{fig:algo:ds}.
The last part in the bottom right corner of $L$ is special and is called the 
\textit{separator} and is colored green.
Parts which would touch entries of \vr{} in the range of the separator perform 
their updates into separate temporary arrays \vtemp{}.
Before the separator is then serially updated, the results of the temporary
arrays are gathered back into \vr{}. 
The backward substitution works the same, just reversed and
only updates to different temporary arrays are not required.
The complete forward substitution and backward substitution  is listed in Algorithm~\ref{alg:algo:fw} and \ref{alg:algo:bw}.

 \begin{algorithm}[tp]
   \begin{algorithmic}[1]
     \Procedure{Backward}{}
       \For{panel $p$ in sep. rev.} \Comment{serial execution}
         \For{\textcolor{blue}{col. $j$ in panel $p$ rev.}} \Comment{unroll}\label{alg:bw:1}
            \State i = \nxindx[p] + offset
            \For{k = \nxlnz[j] + offset; k < \nxlnz[j+1]; ++k}
                \State row = \nindx[i++]
                \State \nr[j] -= \nr[row] \nlnz[k] \Comment{indexed DDOT}
            \EndFor

            \State offset = offset - 1
          \EndFor
        \EndFor
        \For{part in parts} \Comment{parallel execution}
          \For{panel $p$ in part rev.}
            \For{\textcolor{blue}{col. $j$ in panel $p$ rev.}} \Comment{unroll}\label{alg:bw:2}

              \State i = \nxindx[p] + offset

              \For{k = \nxlnz[j] + offset; k < \nxlnz[j+1]; ++k}
                \State row = \nindx[i++]
                \State \nr[j] -=  \nr[row] \nlnz[k] \Comment{indexed DDOT}
              \EndFor

              \State offset = offset - 1

            \EndFor
          \EndFor
        \EndFor
        \EndProcedure
   \end{algorithmic}
   \caption{Backward substitution in PARDISO. Separator (sep.), parts, and
panels are iterated over in reversed (rev.) order.}
   \label{alg:algo:bw}
\end{algorithm}

\section{Application -- Circuit Simulation}
~\label{sec:appl} 

In this section we demonstrate how these developments in sparse direct linear solvers
have advanced integrated circuit simulations.  Integrated circuits are composed 
of interconnected transistors. The interconnects are modeled primarily with 
resistors, capacitors, and inductors. The interconnects route signals through the circuit, 
and also deliver power. Circuit equations arise out of Kirchhoff's current
law, applied at each node, and are generally nonlinear
differential-algebraic equations.  In transient simulation of the
circuit, the differential portion is handled by discretizing the time
derivative of the node charge by an implicit integration formula.  The
associated set of nonlinear equations is handled through use of
quasi-Newton methods or continuation methods, which change the
nonlinear problem into a series of linear algebraic solutions.  Each
component in the circuit contributes only to a few equations.  Hence
the resulting systems of linear algebraic equations are extremely
sparse, and most reliably solved by using direct sparse matrix
techniques.  Circuit simulation matrices are peculiar in the universe
of matrices, having the following characteristics~\cite{davis:klu}:

\begin{itemize}
\item they are unsymmetric, although often nearly structurally
  symmetric;
\item they have a few dense rows and columns (e.g., power and ground
  connections);
\item they are {\em very} sparse and the straightforward usage of
                 BLAS routines (as in SuperLU\cite{superlu}) may 
                 be ineffective;
\item their LU factors remain sparse if well-ordered;
\item they can have high fill-in if ordered with typical strategies;
\item and being unstructured, the highly irregular memory access causes
  factorization to proceed only at a few percent of the peak flop-rate.
\end{itemize}

Circuit simulation matrices also vary from being positive definite to
being {\em extremely} ill-conditioned, making pivoting for stability
important also.  As circuit size increases, and depending on how much
of the interconnect is modeled, sparse matrix factorization is the
dominant cost in the transient analysis.

To overcome the complexity of matrix factorization a new class of
simulators arose in the 1990s, called fast-SPICE \cite{Rewienski2011APO}.
These simulators partition the circuit into subcircuits and use a 
variety of techniques, including model order reduction and multirate
integration, to overcome the matrix
bottleneck.  However, the resulting simulation methods generally incur
unacceptable errors for analog and tightly coupled circuits. As
accuracy demands increase, these techniques become much slower than
traditional SPICE methods. Even so, since much of the research effort
was directed at fast-SPICE simulators, it brought some relief from
impossibly slow simulations when some accuracy trade-off was
acceptable.  Because these simulators partitioned the circuit, and did
not require the simultaneous solution of the entire system of linear
equations at any given time, they did not push the state-of-the-art in
sparse matrix solvers.

Starting in the mid-2000s, increasing demands
on accuracy, due to advancing semiconductor technology, brought
attention back to traditional SPICE techniques.  This was aided by the
proliferation of multicore CPUs. Parallel circuit simulation, an area
of much research focus in the 1980s and 1990s, but not particularly in
practice, received renewed interest as a way to speed up simulation
without sacrificing accuracy.  Along with improved implementations to
avoid cache misses, rearchitecture of code for parallel computing,
and better techniques for exploitation of circuit latency, improved
sparse matrix solvers, most notably the release of KLU
\cite{davis:klu}, played a crucial role in expanding the utility of
SPICE. 

Along with the ability to simulate ever larger circuits with full
SPICE accuracy came the opportunity to further improve sparse matrix
techniques.  A sparse matrix package for transient simulation
needs to have the following features:

\begin{itemize}
\item must be parallel;
\item fast matrix reordering
\item incremental update of the $L$ and $U$ factors when only a few
  nonzeros change;
\item fast computation of the diagonal entries of the inverse matrix;
\item fast computation of Schur-complements for a submatrix;
\item allow for multiple $LU$ factors of the same structure to be stored;
\item use the best-in-class method across the spectrum of sparsity;
\item use iterative solvers with fast construction of sparse preconditioners;
\item run on various hardware platforms (e.g. GPU acceleration).
\end{itemize}

Some of these features must be available in a single package.  Others,
such as iterative solvers and construction of preconditioners, can be
implemented with a combination of different packages. 
The PARDISO solver\footnote{The PARDISO solver is available from 
\url{http://www.pardiso-project.org}.} 
stands out as a package that does most of these very well.
Here we touch on a few of these features.  

When applied in the simulation of very large circuits, the difference between a 
``good'' and a ``bad'' matrix ordering can be the difference between seconds and days.
PARDISO offers AMD and nested-dissection methods for matrix ordering, as well as 
permitting user-defined ordering. Because the matrix re-ordering method which has been 
used most often in circuit simulation is due to Markowitz \cite{markowitz}, and because
modern sparse matrix packages do not include this ordering method, we
briefly describe it here.  The Markowitz method is quite well-adapted for circuit
simulation.  Some desirable aspects of the typical implementation of the Markowitz method,
as opposed to the MD variants, are that it works for
nonsymmetric matrices and combines pivot choice with numerical
decomposition, such that a pivot choice is a numerically ``good''
pivot which generates in a local sense the least fill-in at that step
of the decomposition.  Choosing pivots based on the Markowitz score 
often produces very good results: near-minimal fill-in, unfortunately at the cost of an
$O(n^3)$ algorithm (for dense blocks).  
Even though the Markowitz algorithm has some good properties when applied
to circuit matrices, the complexity of the algorithm has become quite
burdensome.  When SPICE~\cite{nagel:spice2} was originally conceived,
a hundred-node circuit was huge and the Markowitz algorithm was not a
problem.  Now we routinely see netlists with hundreds of thousands of
nodes and postlayout netlists with millions of elements.  As matrix
order and element counts increase, Markowitz reordering time can
become an obstruction. Even as improved implementations of the Markowitz
method have extended its reach, AMD and nested-dissection 
have become the mainstay of simulation of large denser-than-usual matrices.

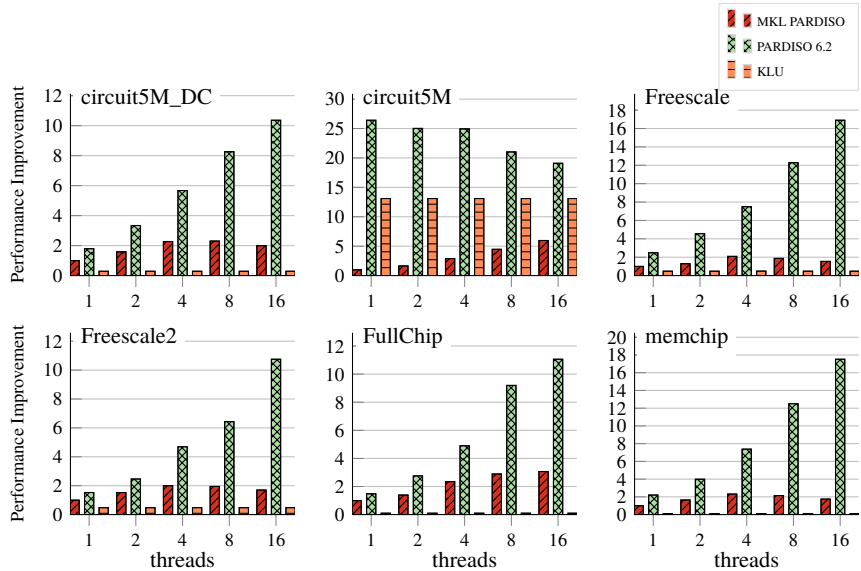
\begin{figure}[t]
\newif\ifrjYLabel
\rjYLabelfalse
\newif\ifrjXTicks
\rjXTicksfalse
\newif\ifrjLegend
\rjLegendfalse
\noindent
\\
\def \rjDataFileName {figures/RJGraphs/circuit5M_DC.dat}
\def \rjTitle {circuit5M\_DC}
\rjYLabeltrue
\begin{tikzpicture}
\pgfplotstableread{\rjDataFileName}\datatable
\begin{axis}[name=symb,
width=0.39\textwidth, height=4cm, 
ybar,
bar width=3.5pt, 
ymode=normal, 
log origin=infty,
ymin=0,
axis lines*=left, 
ymajorgrids, yminorgrids,
xticklabels={1,2,4,8,16},
xtick={1, 2, 3, 4, 5, 6, 7, 8},
xticklabel style={align=center, rotate=0, xshift=-0.0cm, anchor=north, font=\scriptsize},
try min ticks=7,
enlarge y limits={value=0.17,upper},
yticklabel style={font=\scriptsize},
ylabel style={font=\scriptsize},
legend style={at={(1.0,1.50)},fill=white,legend cell align=left,align=right,draw=white!85!black,font=\tiny,legend columns=1},
title=\rjTitle,
every axis title/.style={below right,at={(0,1)},font=\footnotesize,yshift=5pt,xshift=1pt,fill=white}
]
\ifrjYLabel
   \pgfplotsset{ylabel={Performance Improvement}}
\else
\fi
\ifrjXTicks
   \pgfplotsset{xlabel=threads,xlabel style={font=\footnotesize,yshift=4pt}}
\else
   \pgfplotsset{xmajorticks=true}
\fi
\addplot[fill=mycolor1, postaction={pattern=north east lines}] table[x=ID, y=MKL_PARDISO] {\datatable}; 
\ifrjLegend
\addlegendentry{MKL PARDISO}
\fi
\addplot[fill=mycolor3, postaction={pattern=crosshatch}] table[x=ID, y=PARDISO_6_0] {\datatable}; 
\ifrjLegend
\addlegendentry{PARDISO 6.2}
\fi
\addplot[fill=mycolor2, postaction={pattern=horizontal lines}] table[x=ID, y=KLU] {\datatable}; 
\ifrjLegend
\addlegendentry{KLU}
\fi

\end{axis}
\end{tikzpicture}
\def \rjDataFileName {figures/RJGraphs/circuit5M.dat}
\def \rjTitle {circuit5M}
\rjYLabelfalse
\begin{tikzpicture}
\pgfplotstableread{\rjDataFileName}\datatable
\begin{axis}[name=symb,
width=0.39\textwidth, height=4cm, 
ybar,
bar width=3.5pt, 
ymode=normal, 
log origin=infty,
ymin=0,
axis lines*=left, 
ymajorgrids, yminorgrids,
xticklabels={1,2,4,8,16},
xtick={1, 2, 3, 4, 5, 6, 7, 8},
xticklabel style={align=center, rotate=0, xshift=-0.0cm, anchor=north, font=\scriptsize},
try min ticks=7,
enlarge y limits={value=0.17,upper},
yticklabel style={font=\scriptsize},
ylabel style={font=\scriptsize},
legend style={at={(1.0,1.50)},fill=white,legend cell align=left,align=right,draw=white!85!black,font=\tiny,legend columns=1},
title=\rjTitle,
every axis title/.style={below right,at={(0,1)},font=\footnotesize,yshift=5pt,xshift=1pt,fill=white}
]
\ifrjYLabel
   \pgfplotsset{ylabel={Performance Improvement}}
\else
\fi
\ifrjXTicks
   \pgfplotsset{xlabel=threads,xlabel style={font=\footnotesize,yshift=4pt}}
\else
   \pgfplotsset{xmajorticks=true}
\fi
\addplot[fill=mycolor1, postaction={pattern=north east lines}] table[x=ID, y=MKL_PARDISO] {\datatable}; 
\ifrjLegend
\addlegendentry{MKL PARDISO}
\fi
\addplot[fill=mycolor3, postaction={pattern=crosshatch}] table[x=ID, y=PARDISO_6_0] {\datatable}; 
\ifrjLegend
\addlegendentry{PARDISO 6.2}
\fi
\addplot[fill=mycolor2, postaction={pattern=horizontal lines}] table[x=ID, y=KLU] {\datatable}; 
\ifrjLegend
\addlegendentry{KLU}
\fi

\end{axis}
\end{tikzpicture}
\def \rjDataFileName {figures/RJGraphs/Freescale.dat}
\def \rjTitle {Freescale}
\rjLegendtrue
\begin{tikzpicture}
\pgfplotstableread{\rjDataFileName}\datatable
\begin{axis}[name=symb,
width=0.39\textwidth, height=4cm, 
ybar,
bar width=3.5pt, 
ymode=normal, 
log origin=infty,
ymin=0,
axis lines*=left, 
ymajorgrids, yminorgrids,
xticklabels={1,2,4,8,16},
xtick={1, 2, 3, 4, 5, 6, 7, 8},
xticklabel style={align=center, rotate=0, xshift=-0.0cm, anchor=north, font=\scriptsize},
try min ticks=7,
enlarge y limits={value=0.17,upper},
yticklabel style={font=\scriptsize},
ylabel style={font=\scriptsize},
legend style={at={(1.0,1.50)},fill=white,legend cell align=left,align=right,draw=white!85!black,font=\tiny,legend columns=1},
title=\rjTitle,
every axis title/.style={below right,at={(0,1)},font=\footnotesize,yshift=5pt,xshift=1pt,fill=white}
]
\ifrjYLabel
   \pgfplotsset{ylabel={Performance Improvement}}
\else
\fi
\ifrjXTicks
   \pgfplotsset{xlabel=threads,xlabel style={font=\footnotesize,yshift=4pt}}
\else
   \pgfplotsset{xmajorticks=true}
\fi
\addplot[fill=mycolor1, postaction={pattern=north east lines}] table[x=ID, y=MKL_PARDISO] {\datatable}; 
\ifrjLegend
\addlegendentry{MKL PARDISO}
\fi
\addplot[fill=mycolor3, postaction={pattern=crosshatch}] table[x=ID, y=PARDISO_6_0] {\datatable}; 
\ifrjLegend
\addlegendentry{PARDISO 6.2}
\fi
\addplot[fill=mycolor2, postaction={pattern=horizontal lines}] table[x=ID, y=KLU] {\datatable}; 
\ifrjLegend
\addlegendentry{KLU}
\fi

\end{axis}
\end{tikzpicture}
\\
\def \rjDataFileName {figures/RJGraphs/Freescale2.dat}
\def \rjTitle {Freescale2}
\rjYLabeltrue
\rjXTickstrue
\rjLegendfalse
\begin{tikzpicture}
\pgfplotstableread{\rjDataFileName}\datatable
\begin{axis}[name=symb,
width=0.39\textwidth, height=4cm, 
ybar,
bar width=3.5pt, 
ymode=normal, 
log origin=infty,
ymin=0,
axis lines*=left, 
ymajorgrids, yminorgrids,
xticklabels={1,2,4,8,16},
xtick={1, 2, 3, 4, 5, 6, 7, 8},
xticklabel style={align=center, rotate=0, xshift=-0.0cm, anchor=north, font=\scriptsize},
try min ticks=7,
enlarge y limits={value=0.17,upper},
yticklabel style={font=\scriptsize},
ylabel style={font=\scriptsize},
legend style={at={(1.0,1.50)},fill=white,legend cell align=left,align=right,draw=white!85!black,font=\tiny,legend columns=1},
title=\rjTitle,
every axis title/.style={below right,at={(0,1)},font=\footnotesize,yshift=5pt,xshift=1pt,fill=white}
]
\ifrjYLabel
   \pgfplotsset{ylabel={Performance Improvement}}
\else
\fi
\ifrjXTicks
   \pgfplotsset{xlabel=threads,xlabel style={font=\footnotesize,yshift=4pt}}
\else
   \pgfplotsset{xmajorticks=true}
\fi
\addplot[fill=mycolor1, postaction={pattern=north east lines}] table[x=ID, y=MKL_PARDISO] {\datatable}; 
\ifrjLegend
\addlegendentry{MKL PARDISO}
\fi
\addplot[fill=mycolor3, postaction={pattern=crosshatch}] table[x=ID, y=PARDISO_6_0] {\datatable}; 
\ifrjLegend
\addlegendentry{PARDISO 6.2}
\fi
\addplot[fill=mycolor2, postaction={pattern=horizontal lines}] table[x=ID, y=KLU] {\datatable}; 
\ifrjLegend
\addlegendentry{KLU}
\fi

\end{axis}
\end{tikzpicture}
\def \rjDataFileName {figures/RJGraphs/FullChip.dat}
\def \rjTitle {FullChip}
\rjYLabelfalse
\begin{tikzpicture}
\pgfplotstableread{\rjDataFileName}\datatable
\begin{axis}[name=symb,
width=0.39\textwidth, height=4cm, 
ybar,
bar width=3.5pt, 
ymode=normal, 
log origin=infty,
ymin=0,
axis lines*=left, 
ymajorgrids, yminorgrids,
xticklabels={1,2,4,8,16},
xtick={1, 2, 3, 4, 5, 6, 7, 8},
xticklabel style={align=center, rotate=0, xshift=-0.0cm, anchor=north, font=\scriptsize},
try min ticks=7,
enlarge y limits={value=0.17,upper},
yticklabel style={font=\scriptsize},
ylabel style={font=\scriptsize},
legend style={at={(1.0,1.50)},fill=white,legend cell align=left,align=right,draw=white!85!black,font=\tiny,legend columns=1},
title=\rjTitle,
every axis title/.style={below right,at={(0,1)},font=\footnotesize,yshift=5pt,xshift=1pt,fill=white}
]
\ifrjYLabel
   \pgfplotsset{ylabel={Performance Improvement}}
\else
\fi
\ifrjXTicks
   \pgfplotsset{xlabel=threads,xlabel style={font=\footnotesize,yshift=4pt}}
\else
   \pgfplotsset{xmajorticks=true}
\fi
\addplot[fill=mycolor1, postaction={pattern=north east lines}] table[x=ID, y=MKL_PARDISO] {\datatable}; 
\ifrjLegend
\addlegendentry{MKL PARDISO}
\fi
\addplot[fill=mycolor3, postaction={pattern=crosshatch}] table[x=ID, y=PARDISO_6_0] {\datatable}; 
\ifrjLegend
\addlegendentry{PARDISO 6.2}
\fi
\addplot[fill=mycolor2, postaction={pattern=horizontal lines}] table[x=ID, y=KLU] {\datatable}; 
\ifrjLegend
\addlegendentry{KLU}
\fi

\end{axis}
\end{tikzpicture}
\def \rjDataFileName {figures/RJGraphs/memchip.dat}
\def \rjTitle {memchip}
\begin{tikzpicture}
\pgfplotstableread{\rjDataFileName}\datatable
\begin{axis}[name=symb,
width=0.39\textwidth, height=4cm, 
ybar,
bar width=3.5pt, 
ymode=normal, 
log origin=infty,
ymin=0,
axis lines*=left, 
ymajorgrids, yminorgrids,
xticklabels={1,2,4,8,16},
xtick={1, 2, 3, 4, 5, 6, 7, 8},
xticklabel style={align=center, rotate=0, xshift=-0.0cm, anchor=north, font=\scriptsize},
try min ticks=7,
enlarge y limits={value=0.17,upper},
yticklabel style={font=\scriptsize},
ylabel style={font=\scriptsize},
legend style={at={(1.0,1.50)},fill=white,legend cell align=left,align=right,draw=white!85!black,font=\tiny,legend columns=1},
title=\rjTitle,
every axis title/.style={below right,at={(0,1)},font=\footnotesize,yshift=5pt,xshift=1pt,fill=white}
]
\ifrjYLabel
   \pgfplotsset{ylabel={Performance Improvement}}
\else
\fi
\ifrjXTicks
   \pgfplotsset{xlabel=threads,xlabel style={font=\footnotesize,yshift=4pt}}
\else
   \pgfplotsset{xmajorticks=true}
\fi
\addplot[fill=mycolor1, postaction={pattern=north east lines}] table[x=ID, y=MKL_PARDISO] {\datatable}; 
\ifrjLegend
\addlegendentry{MKL PARDISO}
\fi
\addplot[fill=mycolor3, postaction={pattern=crosshatch}] table[x=ID, y=PARDISO_6_0] {\datatable}; 
\ifrjLegend
\addlegendentry{PARDISO 6.2}
\fi
\addplot[fill=mycolor2, postaction={pattern=horizontal lines}] table[x=ID, y=KLU] {\datatable}; 
\ifrjLegend
\addlegendentry{KLU}
\fi

\end{axis}
\end{tikzpicture}
     \label{fig:mklvs62}
\caption{Performance improvements of PARDISO 6.2 against Intel MKL PARDISO for various circuit simulation matrices.}
\end{figure}

Next we turn our attention to parallel performance. 
While KLU remains a benchmark for serial
solvers, for parallel solvers, MKL-PARDISO is often cited as the
benchmark~\cite{Booth2017, Chen2013}.  To give the reader a sense of
the progress in parallel sparse matrix methods, in Figure
\ref{fig:mklvs62} we compare KLU, PARDISO (Version 6.2) to MKL-PARDISO on up to 16 
cores  on an Intel Xeon E7-4880 architecture with 2.5 GHz processors.

Some of the matrices here can be obtained from the SuiteSparse Matrix
Collection, and arise in transistor level full-chip and memory array
simulations. It is clear that implementation of sparse matrix solvers
has improved significantly over the years.  

Exploiting latency in all parts of the SPICE algorithm is very important 
in enabling accurate circuit simulation, especially as the circuit size 
increases.  By latency, we mean that only a few entries in the matrix change from one
\begin{wrapfigure}{r}{0.5\textwidth}
     \centering
     \includegraphics[width=0.5\textwidth]{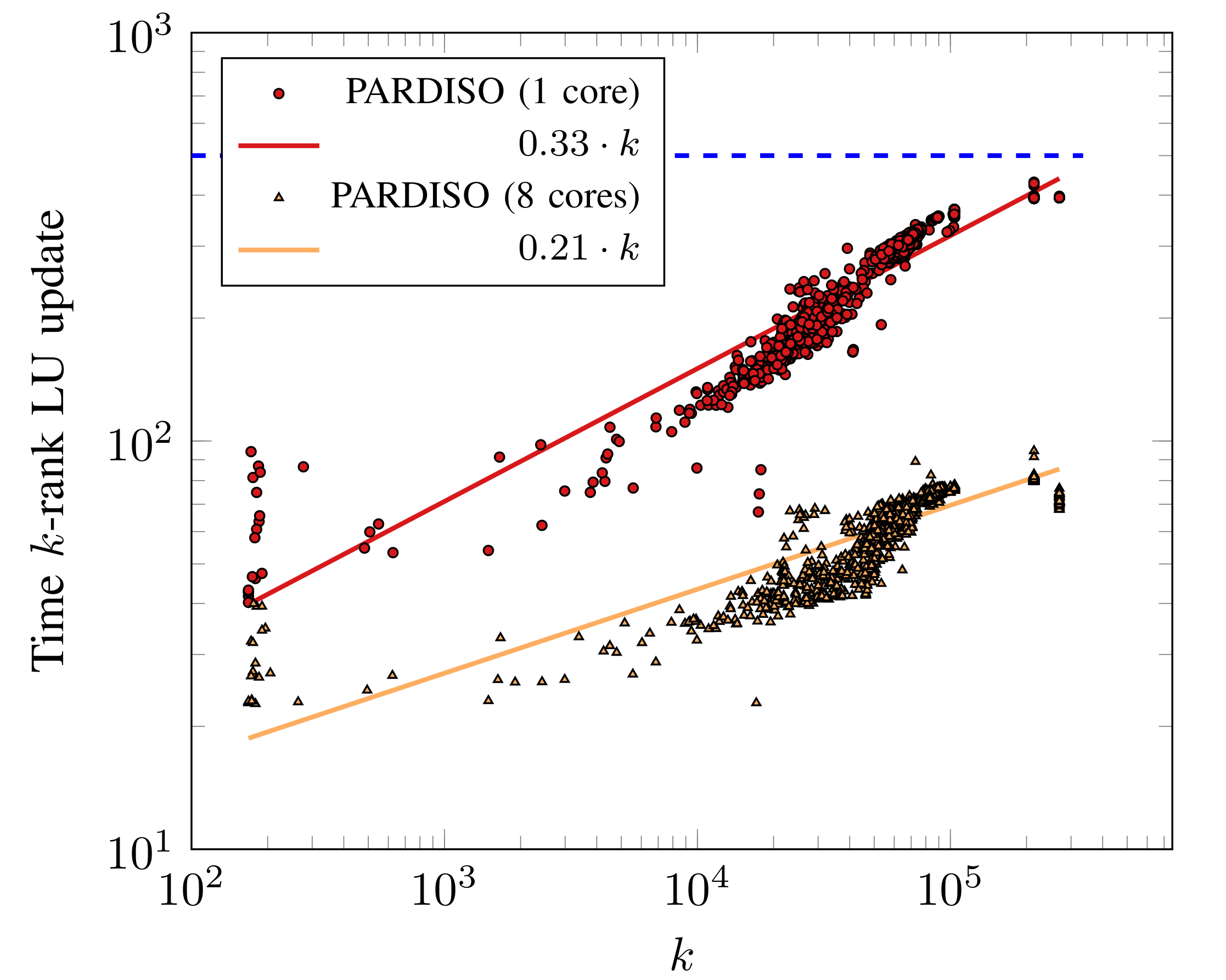} 

     \caption{Regression analysis on the rank-$k$ update $LU$
       factorization in PARDISO. }
     \label{fig:incrementalLU}
\end{wrapfigure}
Newton iteration to the next, and from one timepoint to the next. As
the matrix depends on the time-step, some simulators hold the
time-steps constant as much as feasible to allow increased reuse of
matrix factorizations. The nonzero entries of a matrix change only
when the transistors and other nonlinear devices change their
operation point. In most circuits, very few devices change state from
one iteration to the next and from one time-step to the
next. Nonzeros contributed by entirely linear components do not change
value during the simulation. This makes incremental LU
factorization a very useful feature of any matrix solver used in
circuit simulation.  As of April 2019 the version PARDISO 6.2
 has a very efficient exploitation of
incremental LU factorization, both serial and parallel.   In
Figure~\ref{fig:incrementalLU} we show that PARDISO scales linearly
with number of updated columns, and also scales well with number of
cores. Here, the series of matrices were obtained from a full
simulation of a post-layout circuit that includes all interconnects, 
power- and ground-networks). The factorization time is plotted against the number of
columns that changed compared to the previous factorization.
The scatter plot shows the number of
rank-$k$ update and the corresponding factorization time in
       milliseconds. The regression analysis clearly demonstrates a
       linear trend both for the single and the multiple core
       versions. The dashed line shows the time for the full
       factorization.

Another recent useful feature in PARDISO is parallel selective inverse matrix computation as
demonstrated in Table~\ref{table:bench_matrices}.
In circuit simulation, the diagonal of the inverse matrix is the
driving point impedance. It is often required to flag nodes in the
circuit with very high driving point impedance. Such nodes would
indicate failed interfaces between different subcircuits, leading to
undefined state and high current leakage and power dissipation. A
naive approach to this is to solve for the driving point impedance,
the diagonal of the inverse matrix, by $N$ triangular solves. This is
sometimes unacceptably expensive even with exploiting the sparsity of the
right hand side, and minimizing the number of entries needed in the diagonal of the inverse.
To bypass this complexity, heuristics to compute the impedance of 
connected components are used. But this is error prone with many
false positives and also false negatives. In the circuit Freescale, PARDISO, e.g., finished the
required impedance calculations in 11.9 seconds compared to the
traditional computation that consumed 162.9 hours.

\begin{table}[t]
	\centering
	\caption{Details of the benchmark matrices. 'N' is the number of matrix rows and 'nnz' is the number of nonzeros. The table
                shows the fill-in factor related to the numbers of nonzerors in $\frac{L+U}{A}$, the time for computing all diagonal elements 
                of the inverse $A^{-1}$ using $N$ multiple forward/backward substitution in hours, and  using the selected inverse method in 
		PARDISO for computing all diagonal elements of  the inverse $A^{-1}$ in seconds.}
	\begin{center}
\begin{tabular}{|l|r|r|c|r|r|}\hline
\multicolumn{1}{|c|}{Matrix}  & 
\multicolumn{1}{c|}{N}       &
\multicolumn{1}{c|}{nnz$(A)$}&
\multicolumn{1}{c|}{nnz$(\frac{L+U}{A})$} &  
\multicolumn{1}{c|}{$A^{-1}$} & 
\multicolumn{1}{c|}{Selected $A^{-1}$}  \\\hline
{circuit5M\_DC}	&  3,523,317   & 19,194,193	&  2.87  &  82.3 h.    &  1.3 s.\\
{circuit5M}	&  5,558,326   & 59,524,291	&  1.04  & 371.1 h.    &  2.1 s. \\
{Freescale}	&  3,428,755   & 18,920,347	&  2.94  & 89.8 h.     &  1.0 s. \\
{Freescale2}	&  2,999,349   & 23,042,677	&  2.92  & 8.5  h.      &  1.2 s. \\
{FullChip}	&  2,987,012   & 26,621,990	&  7.41  & 162.9 h.      &  11.9 s. \\
{memchip}	&  2,707,524   & 14,810,202	&  4.40  & 62.5 h.     & 0.9 s. \\\hline

\end{tabular}
 \label{table:bench_matrices}
	\end{center}
\end{table}

The productivity gap in simulation continues to grow, and challenges
remain.  Signoff simulations demand 10X speedup in sparse matrix
factorization. Simply using more cores does not help unless the matrices
are very large and complex. For a majority of simulations, scaling beyond
8 cores is difficult. As a result, some of these
simulations can take a few months to complete, making them essentially
impossible. Some of the problems in parallelizing sparse matrix
operations for circuit simulation are fundamental. Others may be related
to implementation.  Research on sparse matrix factorization for circuit simulation
continues to draw attention, especially in the area of acceleration
with Intel's many integrated core (MIC) architecture \cite{Booth2017}
and GPUs \cite{Chen2015, Nakhla2018}. Other techniques for
acceleration include improved preconditioners for iterative solvers
\cite{Feng2015}. We are presently addressing the need for runtime selection of
optimal strategies for factorization, and also GPU acceleration. Given
that circuits present a wide spectrum of matrices, no matter how we
categorize them, it is possible to obtain a solver that is 2--10X
better on a given problem.  Improvements in parallel sparse matrix
factorization targeted at circuit simulation is more necessary today
than ever and will continue to drive applicability of traditional
SPICE simulation methods.  Availability of sparse matrix packages such
as PARDISO that completely satisfy the needs of various circuit
simulation methods is necessary for continued performance gains.

\bibliographystyle{abbrv}
\begin{small}
\bibliography{direct}
\end{small}
\end{document}